\documentclass[final,authoryear,usenatbib,12pt]{elsarticle}

\usepackage{graphicx}
\usepackage{amsmath}
\usepackage{color}
\usepackage{lineno}

\usepackage{amssymb}

\biboptions{square}

\journal{Astroparticle Physics}

\begin{document}

\begin{frontmatter}



\title{Primordial Black Holes: Observational Characteristics of The Final Evaporation}


\author[LANL]{T. N. Ukwatta}
\author[MSU]{D. R. Stump}
\author[MSU]{J. T. Linnemann}
\author[UNF]{J. H. MacGibbon}
\author[MSU]{S. S. Marinelli}
\author[MSU]{T. Yapici}
\author[MSU]{K. Tollefson}

\address[LANL]{Director's Postdoctoral Fellow, Space and Remote Sensing (ISR-2), Los Alamos National Laboratory, Los Alamos, NM 87545, USA.}
\address[MSU]{Department of Physics and Astronomy, Michigan State University, East Lansing, MI 48824, USA.}
\address[UNF]{Department of Physics, University of North Florida, Jacksonville, FL 32224, USA.}

\begin{abstract}
Many early universe theories predict the creation of Primordial Black Holes (PBHs). PBHs could have masses
ranging from the Planck mass to $10^5$ solar masses or higher depending on the size of the universe at formation. A Black Hole (BH) has a Hawking temperature
which is inversely proportional to its mass. Hence a sufficiently small BH will quasi-thermally radiate
particles at an ever-increasing rate as emission lowers its mass and raises its temperature. The final moments
of this evaporation phase should be explosive and its description is dependent on the particle physics model.
In this work we investigate the final few seconds of BH evaporation, using the Standard Model and
incorporating the most recent Large Hadron Collider (LHC) results, and provide a new parameterization for the instantaneous emission spectrum.
We calculate for the first time energy-dependent PBH burst light curves in the GeV/TeV energy range. Moreover, we explore PBH burst search
methods and potential observational PBH burst signatures. We have found a unique signature in the PBH burst light curves that may be detectable
by GeV/TeV gamma-ray observatories such as the High Altitude Water Cerenkov (HAWC) observatory. The implications of beyond the Standard
Model theories on the PBH burst observational characteristics are also discussed, including potential sensitivity of the instantaneous photon detection rate to a squark threshold in the 5 -10 TeV range.
\end{abstract}

\begin{keyword}
Primordial Black Holes, HAWC, Very High Energy Bursts, Gamma-ray Bursts
\end{keyword}
\end{frontmatter}

\tableofcontents

\section{Introduction}
\label{Intro}

Many current theories of the early universe predict the production of primordial
black holes (PBHs)\cite{Carr2010}. Cosmological density fluctuations and other mechanisms such as those associated with phase transitions in the early universe could have created PBHs with masses of order of, or smaller than, the cosmological horizon size at the time of formation. Depending on the formation mechanism, PBHs could form at
times from the Planck time\footnote{Planck time $\left(\hbar \rm{G}/c^5\right)^{1/2}\simeq 5.391 \times 10^{-44}$ s} to 1 second after
the Big Bang, or later. Hence the initial mass of a PBH could be as
small as the Planck mass\footnote{Planck mass $\left(\hbar c/\rm{G}\right)^{1/2} \simeq 2.176 \times 10^{-8}$ kg} or as massive as $10^5$ solar mass, or higher.

In 1974, Hawking showed by convolving quantum field theory,
thermodynamics and general relativity that a Black Hole\footnote{Throughout this paper, we use the notation `BH' when discussing a black hole irrespective of its formation mechanism or formation epoch and `PBH' when referring to a black hole created in the early universe.} (BH) has a temperature inversely proportional to its
mass and emits photon and particle radiation with thermal spectra~\cite{Hawking1974}. As the BH emits this radiation,
its mass decreases and hence its temperature and flux increase. A PBH that formed with an initial mass of $\sim 5.0 \times 10^{11}$ kg in the early universe should be expiring today~\cite{MacGibbon2008} with a burst of high-energy particles,
including gamma-rays in the MeV to TeV energy range. Thus PBHs are candidate gamma-ray burst (GRB) progenitors~\cite{Halzen1991}.

Confirmed detection of a PBH evaporation event would provide valuable insights into many areas of physics including the early
universe, high energy particle physics and the convolution of gravitation with thermodynamics. Conversely, non-detection of
PBH evaporation events in sky searches would place important limits on models of the early universe.
One of the most important reasons to search for PBHs is to constrain the cosmological density fluctuation
spectrum in the early universe on scales smaller than those constrained by the cosmic microwave background. There is particular
interest in whether PBHs form from the quantum fluctuations associated with many different types of
inflationary scenarios~\cite{Carr2010}. Detection or upper limits on the number density of PBHs can thus inform inflationary models.

PBHs may be detectable by virtue of several effects. For example, PBHs with planetary-scale masses may be detectable by
their gravitational effects in micro-lensing observations~\cite{Griest2011}; or accretion of matter onto PBHs in
relatively dense environments may produce distinct, observable radiation~\cite{Trofimenko1990}. Such situations, however,
should be rare and therefore difficult to use as probes of the cosmological or local PBH distribution. On the other hand, any
PBHs with an initial mass of $\sim 5.0 \times 10^{11}$ kg is expected to explode today in a final burst of Hawking radiation. These events, out to a determinable distance, should be detectable at Earth as sudden bursts of gamma-rays in the sky. 
Numerous observatories have searched for PBH burst events using direct and indirect methods. These methods are sensitive to the PBH distribution
at various distance scales.
Observatories that observe photons or antiprotons at $\sim$100 MeV can probe the cosmologically-averaged or
Galactic-averaged PBH distribution whereas TeV observatories directly probe PBH bursts on parsec scales.
Because it is possible that PBHs may be clustered at various scales, all these searches provide important information.
We also note that the TeV direct search limits~\cite{Alexandreas1993,Amenomori1995,Linton2006,Tesic2012,Glicenstein2013,UkwattaMilagroPBH2013,UkwattaPBH2015a}
apply not only to PBHs but to any nearby presently bursting black holes, regardless of their formation
mechanism or formation epoch, and so equally constrain the number of local bursting BHs which may have formed in the more recent or current universe.
Table~\ref{limit_table} gives a summary of various search methods, the distance scales they probe and their current best limits.

\begin{table}[h]
\begin{center}
\begin{tabular}{|l|c|c|}
\hline Distance Scale & Limit & Method \\ \hline
Cosmological Scale & $< \, 10^{-6}$ ${\rm pc^{-3} yr^{-1}}$ & (1) \\
Galactic Scale & $<$ 0.42 ${\rm pc^{-3} yr^{-1}}$ & (2) \\
Kiloparsec Scale& $<$ $1.2 \times 10^{-3}$ ${\rm pc^{-3} yr^{-1}}$ & (3) \\
Parsec Scale & $<$ $1.4 \times 10^{4}$ ${\rm pc^{-3} yr^{-1}}$ & (4) \\ \hline
\end{tabular}
\caption{PBH burst limits on various distance scales: (1) from the 100 MeV extragalactic gamma-ray background assuming no PBH clustering~\cite{Carr2010, PageHawking1976}, (2)
from the Galactic 100 MeV anisotropy measurement~\cite{Wright1996}, (3) from the Galactic antiproton flux~\cite{Abe2012} and (4) from Very High Energy direct burst searches~\cite{Glicenstein2013}.}
\label{limit_table}
\end{center}
\end{table}

The properties of the BH final burst depend on the physics governing the production and decay of high-energy particles. As the BH evaporates and loses mass over its lifetime, its temperature increases. The higher the number of fundamental particle degrees of freedom, the faster and more powerful will be the final burst from the BH.
The details of the predicted spectra differ according to the high-energy particle physics model.
In the Standard Evaporation Model (SEM) which incorporates the Standard Model of particle physics, a BH should directly Hawking-radiate the fundamental Standard Model particles whose de Broglie wavelengths are of the order of the black hole size~\cite{MacGibbon1990}.
Once the energy of the radiation approaches the Quantum Chromodynamics (QCD) confinement scale ($\sim 200 - 300\  {\rm MeV}$),
quarks and gluons will be directly emitted~\cite{MacGibbon1990}. As they stream away from the BH, the quarks and gluons should fragment and hadronize (analogous to jets seen in
high-energy collisions in terrestrial accelerators) into the particles which are stable on
astrophysical timescales~\cite{MacGibbon2008}. Thus in the SEM, the evaporating black hole is an astronomical burst of photons, neutrinos, electrons, positrons, protons and anti-protons (and for sufficiently nearby sources, neutrons and anti-neutrons~\cite{Smith2013, Keivani2015}).

The purpose of this paper is to examine the observational characteristics of the final evaporation phase of a BH according to the SEM, incorporating the recent Large Hadron Collider (LHC) results for TeV energies, and to explore observational strategies
that can be used in direct PBH burst searches, with particular relevance for the High Altitude Water Cherenkov (HAWC)
observatory. Included is a discussion of the limitations and advantages of specific burst search methods and unique PBH burst signatures.
In Section~\ref{sec:pbh_theory}, we review the black hole Hawking radiation process. In Section~\ref{sec:burstphotons}, we
use an empirical fragmentation function to calculate the BH photon spectrum and light curve and to parameterize the instantaneous photon emission from a BH burst.
In Section~\ref{sec:pbh_search}, we explore methods for direct searches for PBH bursts and the procedures for setting upper limits on the PBH distribution which would arise from null detection. We also discuss how one
can potentially differentiate a PBH burst from other known cosmological GRB sources. In Section~\ref{sec:BSM}, modifications that could arise from high energy physics beyond the Standard Model are elucidated. In Section~\ref{sec:discussion}, we discuss the applicability and limitations of various assumptions employed in our PBH burst properties calculations.
A summary of our findings and conclusions is given in Section~\ref{sec:conclusion}.

\clearpage

\section{BH Emission Theory}\label{sec:pbh_theory}

\subsection{Hawking Radiation}\label{sec:Hawking_rad}

Hawking showed that a black hole radiates each fundamental particle species at
an emission rate of~\cite{Hawking1974, Hawking1975}
\begin{equation}\label{eq:hawking}
\frac{d^{2}N}{dE dt} = \frac{\Gamma/2\pi\hbar}{e^{x} - (-1)^{2s}} \ n_{\rm dof} ,
\end{equation}
where $s$ is the particle spin, $n_{\rm dof}$ is the number of degrees of freedom of the
particle species (e.g. spin, electric charge, flavor and color), $\Gamma$ is the absorption coefficient, and $\hbar$ is the reduced Planck constant. The dimensionless quantity $x$ is defined by
\begin{equation}\label{eq:xeq}
x \equiv \frac{8\pi G M_{BH} E}{\hbar c^{3}} = \frac{E}{kT_{BH}}
\end{equation}
for a nonrotating, uncharged 4D black hole, where $E$ is the energy of the Hawking-radiated particle, $M_{BH}$ is the black hole mass, $T_{BH}$ is the black hole temperature,
\begin{equation}\label{eq:tempmass}
kT_{BH} = \frac{\hbar c^{3}}{8\pi G M_{BH}} = 1.058\left(\frac{10^{10}\ \rm{kg}}{M_{BH}}\right)\ \rm{GeV},
\end{equation}
$G$ is the universal gravitational constant, $c$ is the speed of light and $k$ is Boltzmann's constant. Because initial black hole rotation and/or electric charge is radiated away faster than mass, we will assume a nonrotating, uncharged black hole in our analysis; extension to rotating and/or charged black holes is straightforward ~\cite{MacGibbon1990, Page1976b, Page1977}.

\begin{figure}\begin{center}
\includegraphics[width=0.9\textwidth]{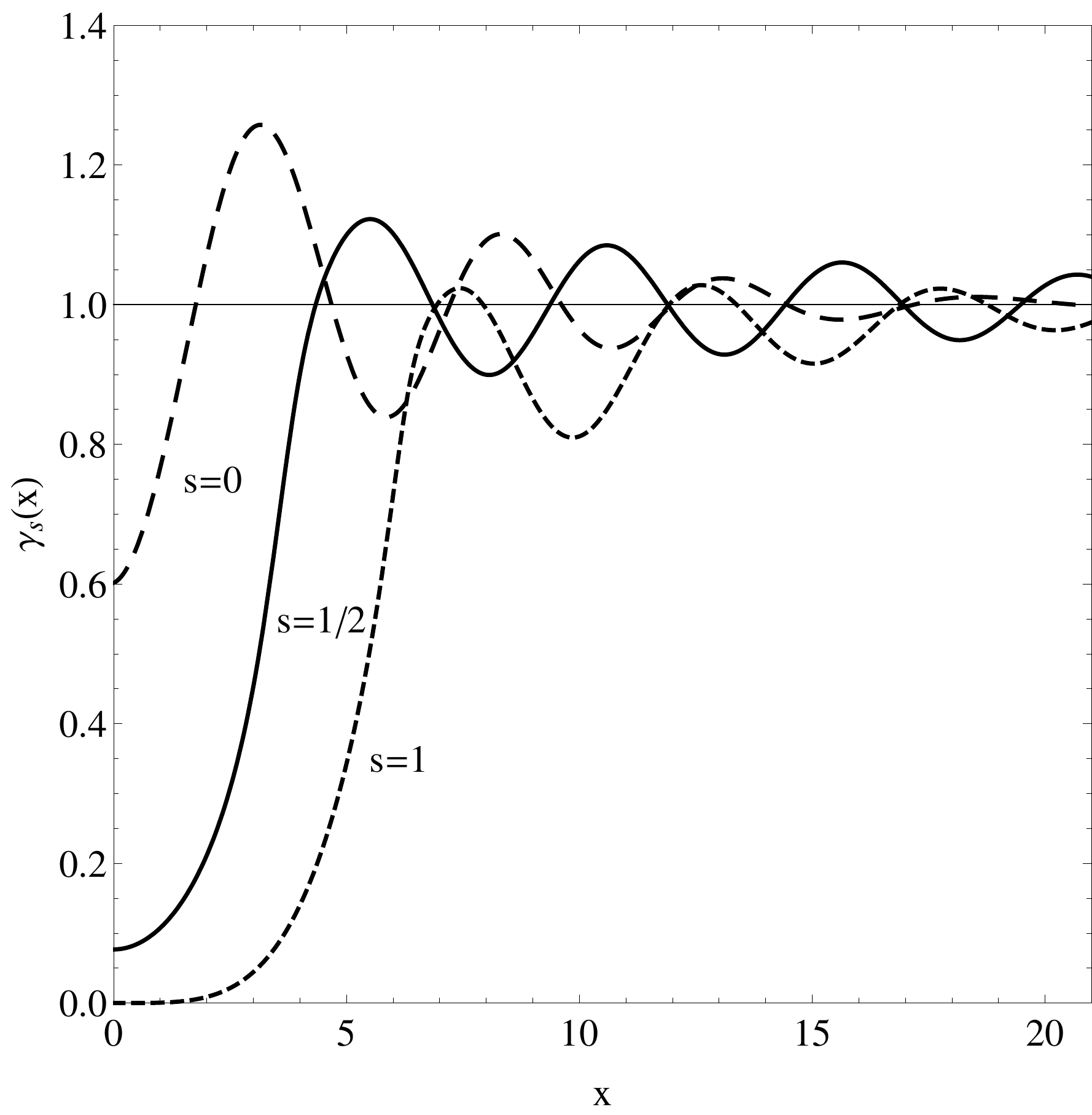}
\caption{The functions $\gamma_{s}(x)$
for massless spin-0 (long dashed line), spin-1/2 (solid line) and spin-1 (short dashed line) particles.
\label{fig:PageFunc}}\end{center}
\end{figure}

The absorption coefficient $\Gamma$ depends on $M_{BH}$, $E$ and $s$. For an emitted species of rest
mass $m$, $\Gamma$ at $E \gg mc^{2}$ has the form
\begin{equation}\label{eq:Page}
\Gamma(M_{BH}, E, s) = 27 \left(\frac{x}{8\pi}\right)^{2} \gamma_{s}(x)
\end{equation}
such that $\gamma_{s}(x)\rightarrow 1$ for large $x$.
The functions $\gamma_{s}(x)$ are shown in Fig.~\ref{fig:PageFunc}
for massless or relativistic uncharged particles with $s = 0$, $s = 1/2$ and $s = 1$
\cite{MacGibbon1990, Page1976a, Page1976b, Page1977, Elster1983a, Elster1983b, Simkins1986}. For a non-relativistic $s=1/2$ particle, $\Gamma$ at $E= mc^2$ remains at least $50\%$ of the relativistic value and, when $kT_{BH}\gtrsim 0.1 mc^2$, only deviates noticeably from the relativistic value at $E \lesssim 2 mc^2$ \cite{Page1977}. Below $E= mc^2$, $\Gamma = 0$. Electrostatic effects associated with the emission of a particle of electric charge $e$ decrease $\Gamma$ by at most a few percent\cite{Page1977}.

\begin{figure}\begin{center}
\includegraphics[width=0.9\textwidth]{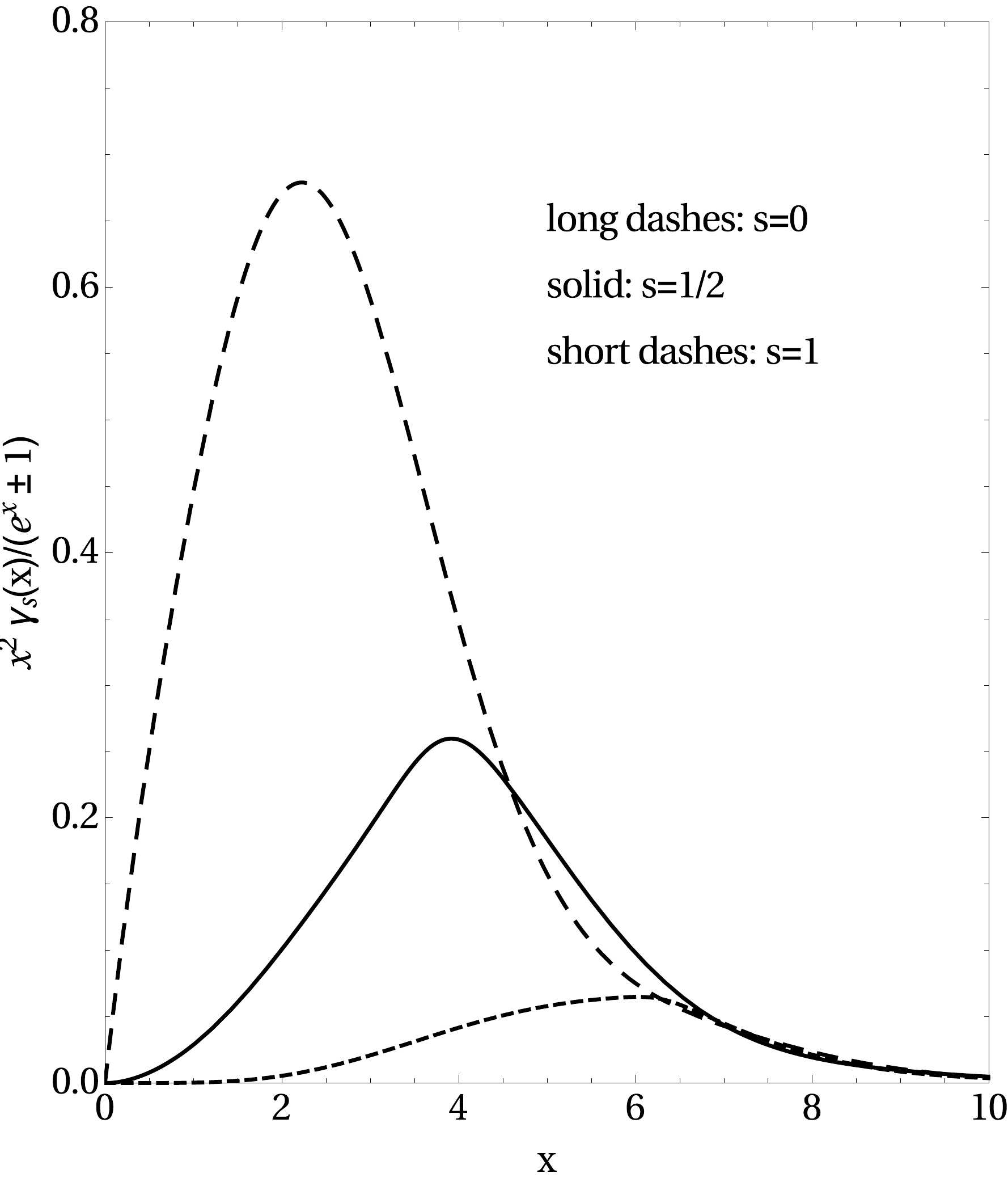}
\caption{The dimensionless emission rate per degree of freedom, $\psi_s(x) = x^2 \gamma(x)/(e^x \pm 1)$ as a function of $x$ for $s=0$, $1/2$, and 1.
\label{fig:PageFunc1}}\end{center}
\end{figure}

Combining the above equations, the emission rate per fundamental particle species can be written in the form
\begin{equation}\label{eq:psiflux}
\frac{d^{2}N}{dE dt} = \frac{27}{2\pi\hbar(8\pi)^2} n_{\rm dof} \psi_s (x)
\end{equation}
where
\begin{equation}\label{eq:psi}
\psi_s (x) \equiv \frac{\gamma_s (x) x^2}{e^{x} - (-1)^{2s}}.
\end{equation}
The dimensionless emission rate functions, $\psi_s(x)$, are plotted in Fig.~\ref{fig:PageFunc1} for $s=0$, $1/2$, and $0$. The distribution $\psi_s (x)$ peaks at $x=x_{r,s}$ where $x_{r,s=0} = 2.19$ for uncharged massless or relativistic particles, $x_{r,s=1/2} = 4.02$ for relativistic particles with charge $\pm e$, and $x_{r,s=1} = 5.77$ for uncharged massless or relativistic particles~\cite{MacGibbon1990, Page1976a, Page1977, Elster1983a, Elster1983b, Simkins1986}. The emission rate integrated over energy, per emitted fundamental species, is 
\begin{eqnarray}\label{eq:rate}
\frac{dN}{dt} & = & \int_0^\infty \frac{d^2 N}{dE dt} dE \\
& = & \frac{27 c^3}{2\pi G(8\pi)^3 M_{BH}} n_{\rm dof} \Psi_s = \frac{1.093\times 10^{22}}{\left(M_{BH} / 10^{10}\ \rm{kg}\right)} n_{\rm dof} \Psi_s \ \rm{s}^{-1}\\
& = & \frac{27 k T_{BH}}{2\pi \hbar(8\pi)^2} n_{\rm dof} \Psi_s = 1.033\times 10^{22} \left( \frac{T_{BH}}{\rm{GeV}}\right) n_{\rm dof} \Psi_s \ \rm{s}^{-1}
\end{eqnarray}
where
\begin{equation}\label{eq:Psi}
\Psi_s \equiv \int_0^\infty \frac{\gamma_s (x) x^2}{e^{x} - (-1)^{2s}} dx.
\end{equation}
Per degree of freedom, $\Psi_{s=0} = 2.45$ for uncharged massless or relativistic particles, $\Psi_{s=1/2} = 0.897$ for uncharged relativistic particles, $\Psi_{s=1/2} = 0.879$ for relativistic particles with charge $\pm e$, and $\Psi_{s=1} = 0.273$ for uncharged massless or relativistic particles~\cite{MacGibbon1990, Page1976a, Page1977, Elster1983a, Elster1983b, Simkins1986}.

\begin{figure}\begin{center}
\includegraphics[width=0.9\textwidth]{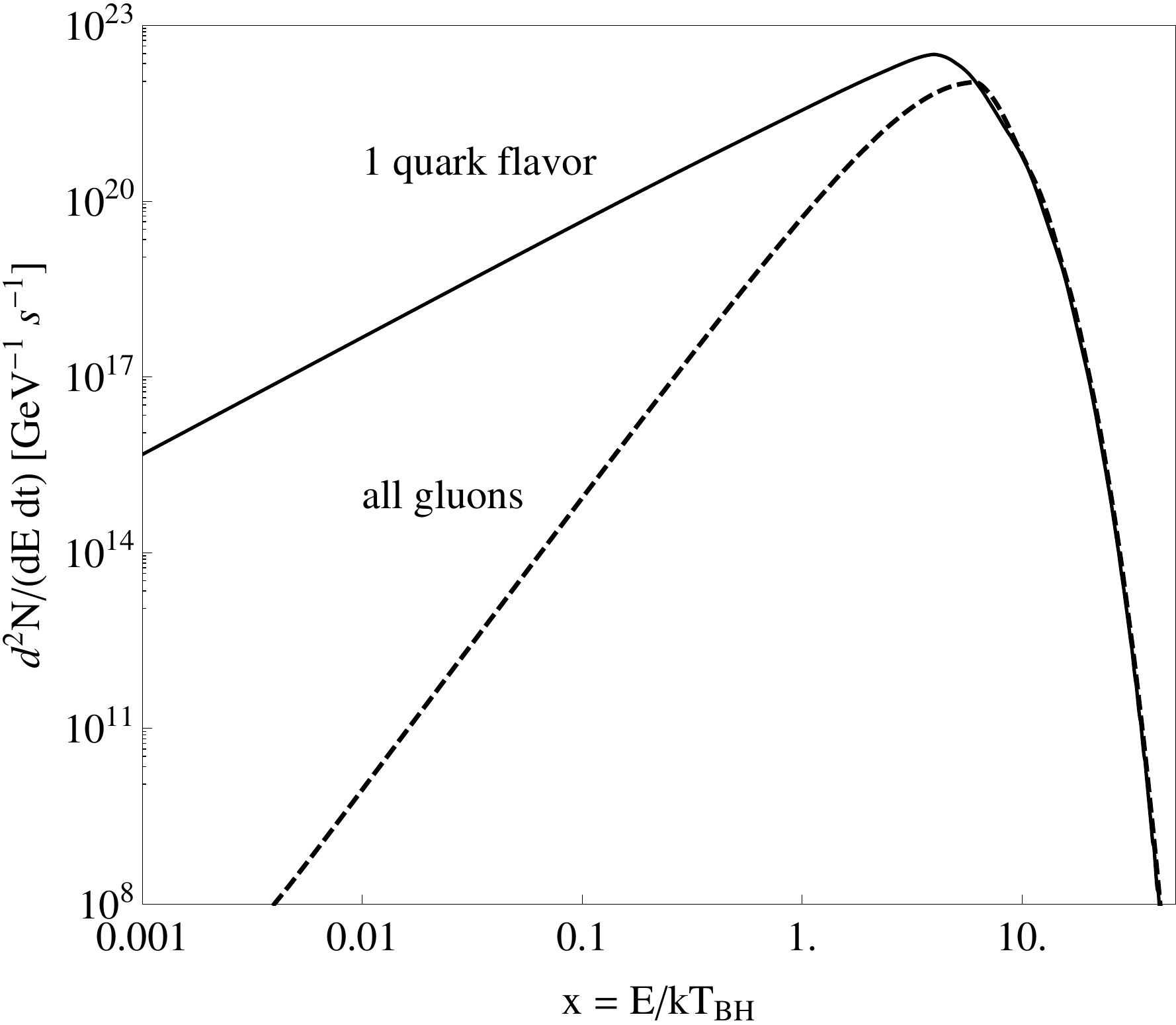}
\caption{
Rates of Hawking radiation of a massless quark flavor and of gluons, as functions of $x$.
For $x << 1$, the power laws are $\psi \propto x^2$ for $s=1/2$ and $\psi \propto x^{3}$ for $s=1$.
\label{fig:primaryfluxes}}\end{center}
\end{figure}

Fig.~\ref{fig:primaryfluxes} displays the direct radiation rates according
to Eq.~\ref{eq:hawking} for a single relativistic quark flavor ($n_{\rm dof} = 12$)
and for gluons ($n_{\rm dof} = 16$), as functions of $x$. (In Fig.~\ref{fig:primaryfluxes}
we have neglected the electric charge of the quark which affects the quark emission rate by less than $5$ percent~\cite{MacGibbon1990, Page1977}.)

In order to calculate the spectrum of the final photon burst from the PBH, two important relations pertaining to the final phase of BH evaporation are needed.
The first relation we require is the black hole mass $M_{BH}$ as a function of time~\cite{Page1976a}
\begin{equation}\label{eq:massloss}
\frac{dM_{BH}}{dt}\equiv -\frac{\alpha(M_{BH})}{M_{BH}^{2}},
\end{equation}
where the function $\alpha(M_{BH})$ incorporates all directly emitted particle species and their degrees of freedom. As the BH evaporates, the value of $M_{BH}$ is reduced by an amount equal to the total mass-energy of the emitted particles.
By conservation of energy,
\begin{equation}\label{disc:conserve_energy}
\frac{d(M_{BH}c^2)}{dt} = - \sum_i \int_0^\infty \frac{d^2 N_i}{dEdt} E dE
\end{equation}
where the summation over $i$ is over all the fundamental species and so
\begin{equation}\label{disc:alpha2}
\alpha(M_{BH}) = \frac{M_{BH}^2}{c^2} \sum_i \int_0^\infty \frac{d^2 N_i}{dEdt} E dE.
\end{equation}
Substituting for $M_{BH}$ in terms of $T_{BH}$ and $E$ in terms of $x$, we can write
\begin{equation}\label{disc:alpha3}
\alpha(M_{BH}) = \frac{27\hbar c^4}{2 \pi G^2 (8 \pi)^4} \sum_i \int_0^\infty n_{{\rm dof},i} \phi_{s_i} (x) dx
\end{equation}
where $\phi_s (x) \equiv \psi_s (x) x$.

\begin{figure}\begin{center}
\includegraphics[width=0.9\textwidth]{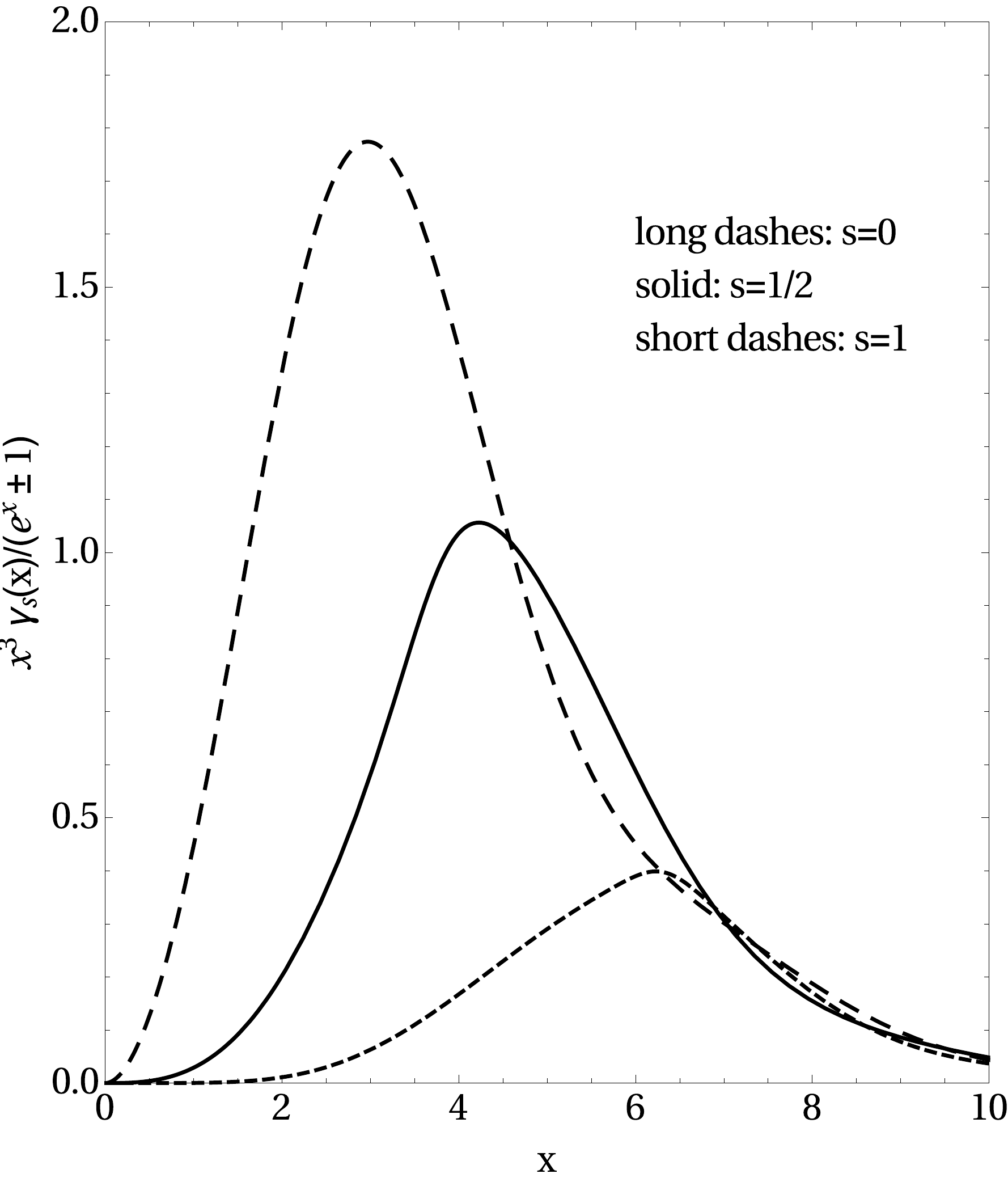}
\caption{The dimensionless emitted power per degree of freedom, $\phi_s(x) = x^3 \gamma(x)/(e^x \pm 1)$ as a function of $x$ for $s=0$, $1/2$, and $1$.
\label{fig:PageFunc2}}\end{center}
\end{figure}

The dimensionless emitted power functions, $\phi_s(x)$, are shown in Fig.~\ref{fig:PageFunc2} for $s=0$, $1/2$, and $0$. The distribution $\phi_s (x)$, and hence the instantaneous power emitted in each fundamental state, peaks at $x=x_{p,s}$ where $x_{p,s=0} = 2.66$ for uncharged massless or relativistic particles, $x_{p,s=1/2} = 4.40$ for uncharged relativistic particles, $x_{p,s=1/2} = 4.53$ for relativistic particles with charge $\pm e$, and $x_{p,s=1} = 6.04$ uncharged massless or relativistic particles~\cite{MacGibbon1990, Page1976a, Page1977, Elster1983a, Elster1983b, Simkins1986}. As the remaining BH evaporation lifetime $\tau$ decreases, $T_{BH}$ increases and new fundamental quanta begin to contribute significantly to $\alpha (M_{BH})$ once $T_{BH}$ crosses
each relevant mass threshold, $kT_{BH} \sim m_i c^2/x_{p,s_i}$. At $kT_{BH} \gtrsim m_i c^2$, the contribution of a specific fundamental species $i$ to $\alpha (M_{BH})$ is
\begin{equation}\label{disc:ai}
\alpha_i = \frac{27\hbar c^4}{2 \pi G^2 (8 \pi)^4} n_{{\rm dof,}i} \Phi_{s_i}
= 2.06 \times 10^{15} n_{{\rm dof,}i} \Phi_{s_i} \rm \,\,\, kg^3 s^{-1},
\end{equation}
where
\begin{equation}\label{disc:Phi}
\Phi_s = \int_0^\infty \frac{\gamma_s(x) x^3}{e^x - (-1)^{2s}} dx.
\end{equation}
Per degree of freedom, $\Phi_{s=0} = 6.89$ for massless or relativistic particles, $\Phi_{s=1/2} = 3.79$ for uncharged relativistic particles, $\Phi_{s=1/2} = 3.68$ for relativistic particles with charge $\pm e$, and $\Phi_{s=1} = 1.56$ for massless or relativistic particles~\cite{MacGibbon1990, Page1976a, Page1977, Elster1983a, Elster1983b, Simkins1986}. From the values for uncharged and $\pm e$-charged $s=1/2$ modes, we can linearly interpolate approximate values for relativistic $\pm e/3$-charged $d$, $s$ and $b$ quarks ($\Phi_{s=1/2} \simeq 3.75$ per {\it d}, {\it s} or {\it b} degree of freedom) and for relativistic $\pm 2e/3$-charged $u$, $c$ and $t$ quarks ($\Phi_{s=1/2} \simeq 3.72$ per {\it u}, {\it c} or {\it t} degree of freedom).

Counting only the experimentally confirmed fundamental Standard Model particles~\cite{Beringer2012}, a $kT_{BH} \simeq 1$ GeV ($M_{BH}\simeq 10^{10}$ kg) black hole should directly emit the following field quanta:
the three charged leptons $(s=1/2,\ n_{dof}=12)$;
the three neutrinos $(s=1/2,\ n_{dof}=6)$ where we assume that the neutrinos
are Majorana particles with negligible mass;
five quark flavors $(s=1/2,\ n_{dof}=60)$;
the photon $(s=1,\ n_{dof}=2)$; and
the gluons $(s=1,\ n_{dof}=16)$. This gives
\begin{equation}
\alpha(kT_{BH} \simeq 1\,\rm GeV)= 6.6 \times 10^{17}
~{\rm kg}^{3}{\rm s}^{-1}.
\end{equation}

At $kT_{BH}\simeq 50$ GeV ($M_{BH}\simeq 2\times 10^{8}$ kg), the list also includes
the top quark, the $W^{\pm}$ and $Z^{0}$ massive vector bosons ($s=1,\ n_{dof}=9$),
and the Higgs boson ($s=0,\ n_{dof}=1$, and
treating the Higgs boson as a 125 GeV resonance~\cite{Aad2015}). This gives
\begin{equation}
\alpha(kT_{BH} \simeq 50\,\rm GeV)= 8.0 \times 10^{17}
~{\rm kg}^{3}{\rm s}^{-1}.
\end{equation}

Energies well above the Higgs field vacuum expectation value $\sim 246$ GeV have not yet been explored in high energy accelerators. The ten fundamental modes of $W^{\pm}, Z^{0}$ and $H^{0}$
are expected to be counted differently above the electroweak symmetry breaking phase transition because of expected restoration of SU(2)xU(1) gauge
symmetry~\cite{Chanowitz1988}, although this has yet to be confirmed in accelerator experiments. By the Goldstone Boson Equivalence Theorem, the longitudinal
modes of the $W^\pm$ and $Z^0$ bosons observed at lower energies are expected to be expressed as scalar modes
at these energies. In this case, there would be 6 transverse vector $s=1$ fields and 4 scalar $s=0$ fields of $W^{\pm}, Z^{0}$ and $H^{0}$, giving an the
asymptotic value of $\alpha(M_{BH})\simeq 8.3\times 10^{17}\ \rm{kg}^3\rm{s}^{-1}$ for $kT_{BH}> 100$ GeV ($M_{BH} < 10^{8}$ kg). Because this has not yet been
observed experimentally and there are other possible arrangements at high energies, however, we will confine our modes to those which have been experimentally
confirmed and use as our asymptotic value of
\begin{equation}
\alpha_{SM} = 8.0 \times 10^{17}
~{\rm kg}^{3}{\rm s}^{-1}
\end{equation}
in subsequent calculations for $kT_{BH}> 100$ GeV
($M_{BH} < 10^{8}$ kg). We note that the ambiguity in counting $s=0$ and $s=1$ states as $T_{BH}$ transitions
through and above the electroweak symmetry breaking scale has a negligible effect on the BH emission
spectra because of the dominance of the $s=1/2$ modes at these $T_{BH}$.

\begin{figure}
\begin{center}
\includegraphics[width=0.85\textwidth]{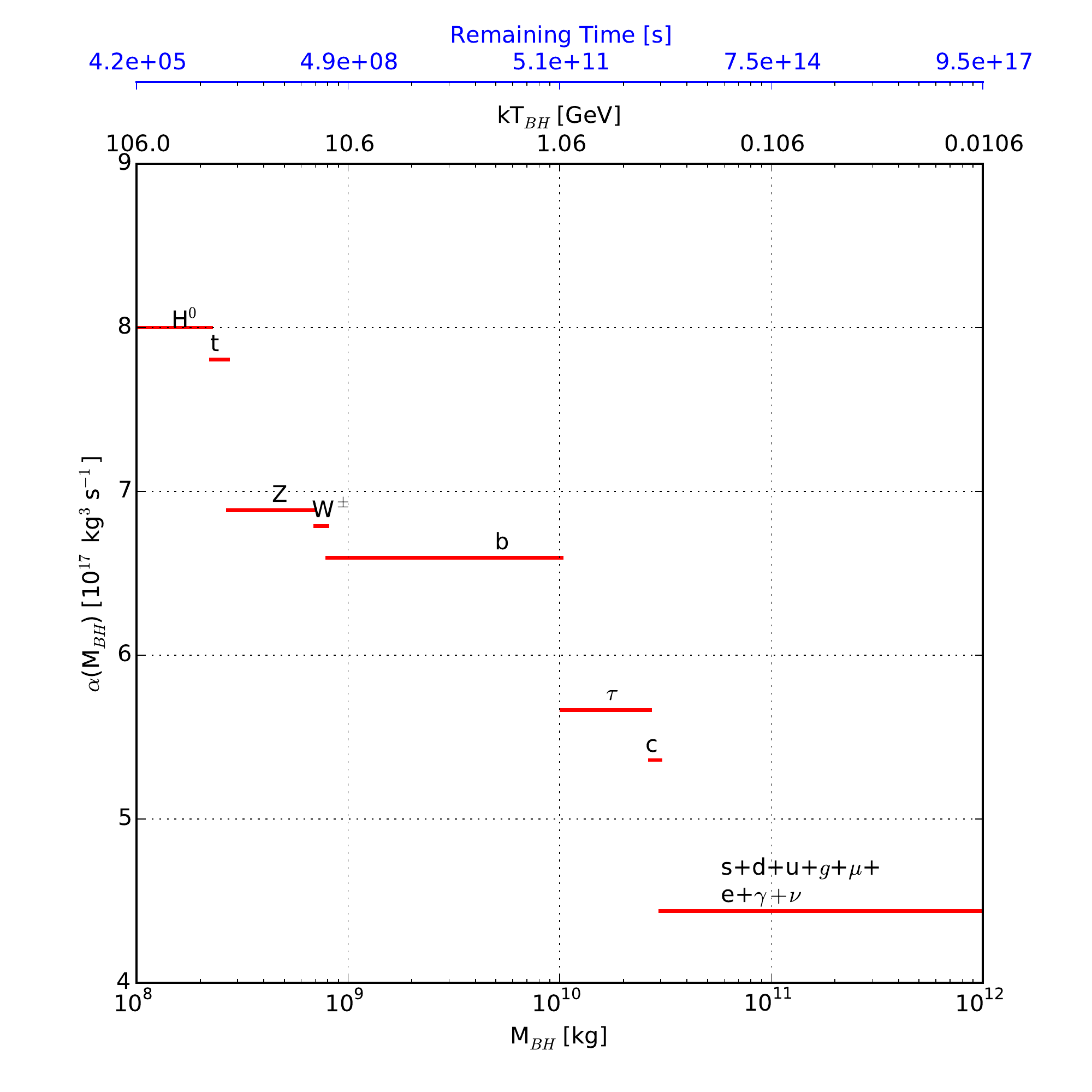}
\caption{The function $\alpha(M_{BH})$ in the Standard Evaporation Model.
The asymptotic value for the SEM is $8.0 \times 10^{17} \ {\rm kg}^{3}{\rm s}^{-1}$.
\label{fig:alphaSEM}}
\end{center}
\end{figure}

Fig.~\ref{fig:alphaSEM} illustrates $\alpha(M_{BH})$ and $\alpha(T_{BH})$ for the SEM. In Fig.~\ref{fig:alphaSEM}, the function is presented as piece-wise constant with each horizontal line segment representing
the sum of the asymptotic contributions for $kT_{BH} > m_{i} c^{2}/x_{p, s_i}$.
As the end of the BH's lifetime approaches,
$T_{BH}$ exceeds the rest masses of all known fundamental particles,
and $\alpha$ reaches a constant asymptotic value, $\alpha_{SM}$.

For the current and future generations of very high energy (VHE) gamma-ray observatories,
we are interested in bursts generated by black holes of
temperature $kT_{BH}\gtrsim 1\, {\rm TeV}$. For $kT_{BH} \gtrsim 1\, {\rm TeV}$ 
(corresponding to $M_{BH} \lesssim 10^{7} \ {\rm kg}$ and a remaining
evaporation lifetime of $\tau \lesssim 500\,{\rm s}$), we have 
$\alpha(M_{BH}) \approx \alpha_{\rm SM}$.
Returning to Eq.~\ref{eq:massloss}, the BH mass as a function of remaining evaporation lifetime $\tau$ in this regime is then
\begin{equation}\label{eq:masstau}
M_{BH}(\tau) \approx \left(3\alpha_{\rm SM}\, \tau\right)^{1/3} =
1.3 \times 10^{6} \left(\frac{\tau}{1 s}\right)^{1/3}\ {\rm kg}.
\end{equation}

The second relation we require is the BH temperature $T_{BH}$
expressed as a function of $\tau$ for the final evaporation phase. 
Combining Eq.~\ref{eq:tempmass} and Eq.~\ref{eq:masstau}, 
we have for $kT_{BH}\gtrsim 1\, {\rm TeV}$
\begin{equation}\label{eq:temptau}
kT_{BH} = 7.8 \left(\frac{\tau}{1 s}\right)^{-1/3}\ {\rm TeV}
\end{equation}
\begin{equation}\label{eq:tautemp}
\tau = 4.8\times 10^{2} \left(\frac{kT_{BH}}{\rm{TeV}}\right)^{-3}\ \rm{s}.
\end{equation}

Strictly, the above equations apply provided that the black hole temperature
is below the Planck temperature $T_{\rm Pl}$ ($\simeq 1.22 \times 10^{16}$ TeV).
However, because the remaining evaporation lifetime dramatically
shortens as $T_{BH}$ increases, the behavior of the BH close to $T_{\rm Pl}$ has negligible
effect on the astronomically observable emission spectra.

{\begin{figure}\begin{center}
\includegraphics[width=0.95\textwidth]{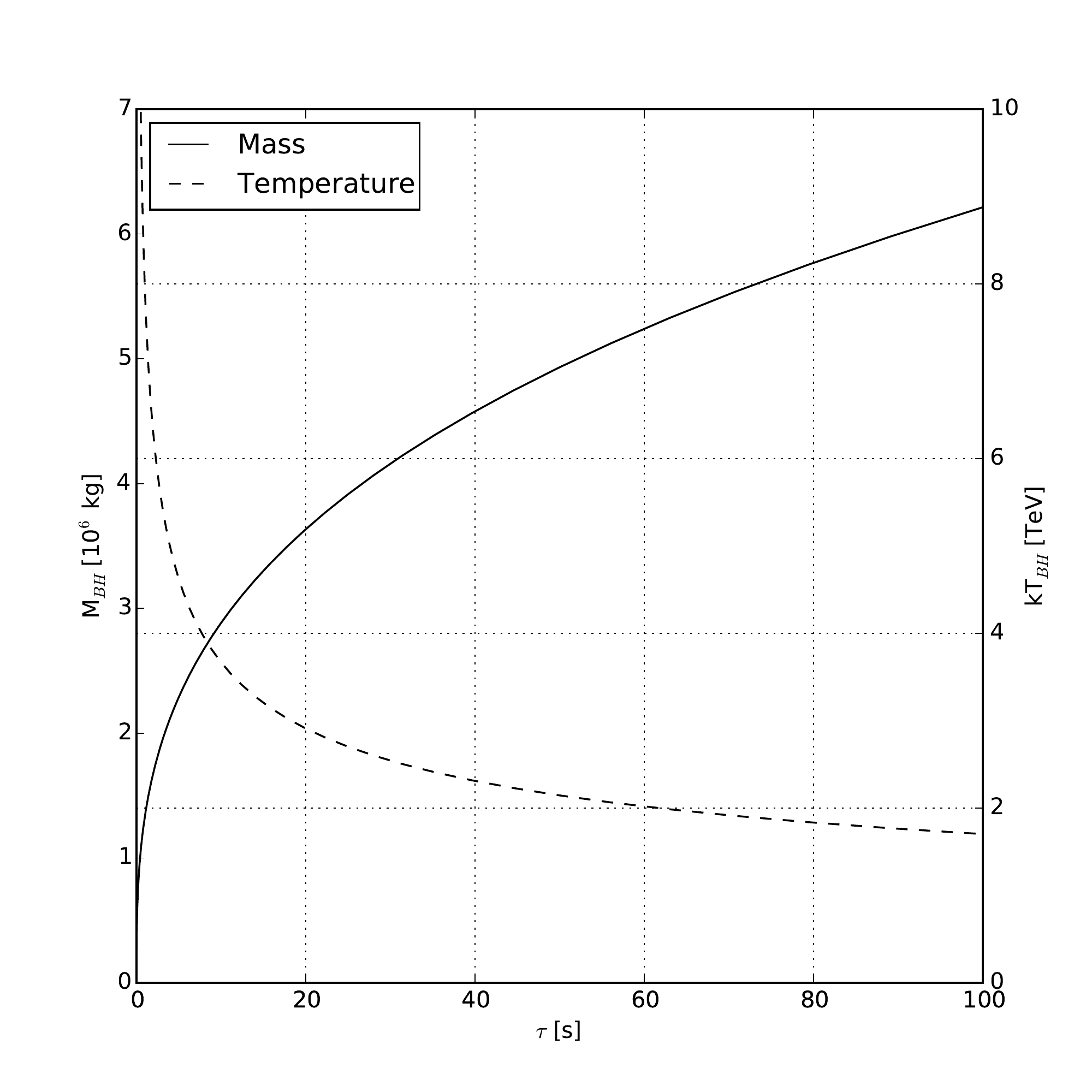}
\caption{Black hole mass and temperature for the final 100 seconds of the BH evaporation lifetime ($\tau$ is the remaining time). The decrease of mass and the increase of temperature accelerate as $\tau \rightarrow$ 0.\label{fig:MandT}}
\end{center}
\end{figure} }
The black hole mass and temperature for the final $100\ \rm{s}$ of evaporation lifetime, corresponding to temperatures $kT_{BH} \gtrsim 2\ {\rm TeV}$, are shown in Fig.~\ref{fig:MandT}. VHE gamma-ray observatories are sensitive to photon energies in the range from $\sim 50$ GeV to 100 TeV. Thus the relevant $x$ range for the final $100\ \rm{s}$ of the PBH burst is $0 \lesssim x \lesssim 50$. The instantaneous emission
rates for a relativistic quark flavor and for gluons as a function of $x$ in this range is included in Fig.~\ref{fig:primaryfluxes}.

In order to elucidate the behaviour near the end of the black hole's evaporation lifetime,
we now investigate the emission rate and spectrum as functions of $\tau$.
Fig.~\ref{fig:quarkflux} shows the instantaneous emission rate
$d^{2}N/(dE dt)$ for a relativistic or massless quark flavor, as a function of quark energy $E$,
when $\tau = 100,\ 10,\ 1,\ 0.1,\ 0.01$ and $0.001$ s.
As $\tau\rightarrow 0$ and $T_{BH}$ increases, the emission rate per directly
Hawking-radiated $s=1/2$ degree of freedom is constant at the peak but increases at high energies and decreases at low energies because the location of the peak scales with $T_{BH}$.

{\begin{figure}\begin{center}
\includegraphics[width=0.75\textwidth]{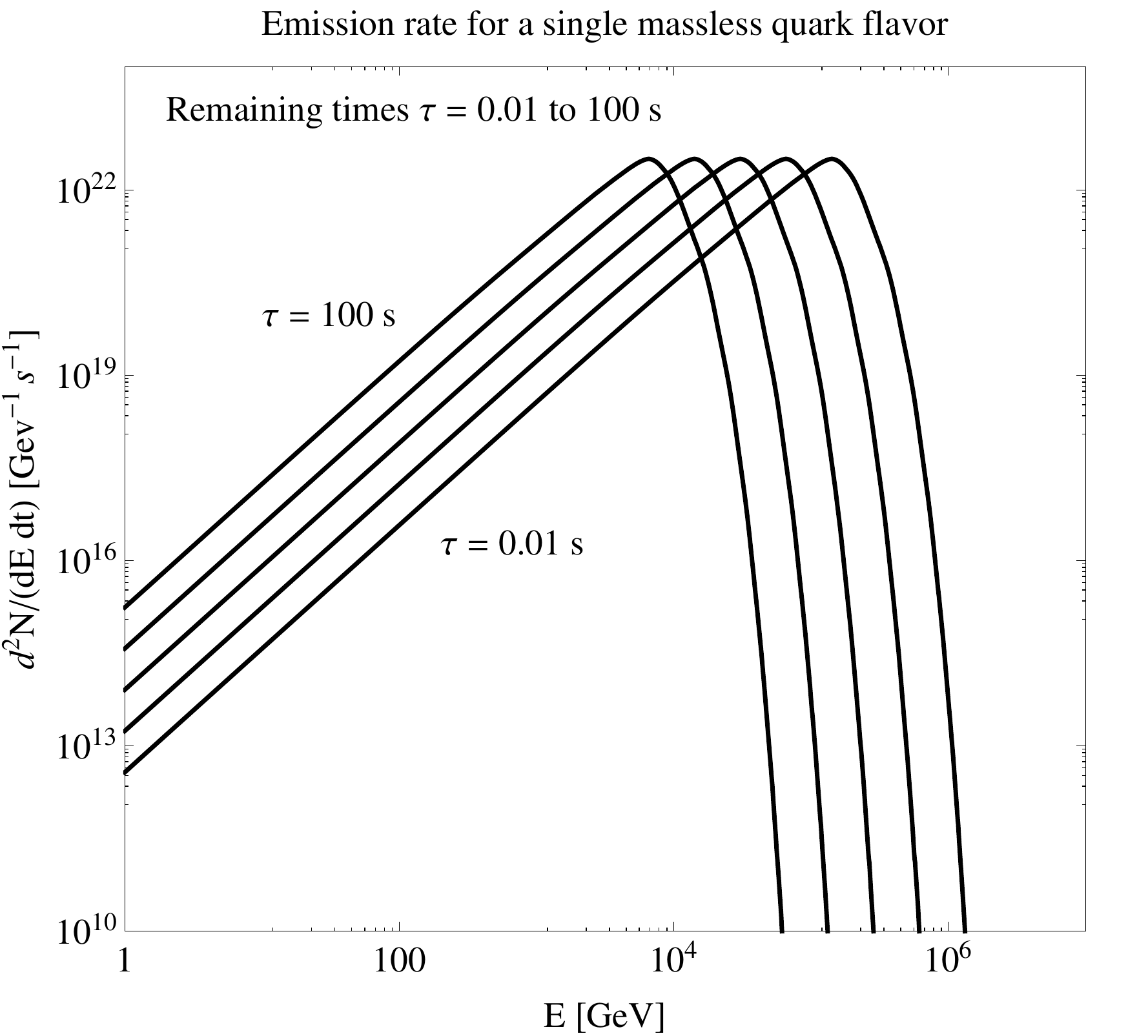}
\caption{Instantaneous number emission rate for a single massless quark flavor,
as a function of quark energy $E$, at five values of
remaining evaporation lifetime $\tau$:
$100,\ 10,\ 1,\ 0.1,\ 0.01$ seconds.
\label{fig:quarkflux}}
\end{center}\end{figure}}

The SEM theory of Hawking radiation
in the final 100 seconds has only one parameter with physical dimensions,
which may be taken to be $kT_{BH}$.
Because $kT_{BH} \gtrsim 2\ {\rm TeV}$ is much larger than any Standard Model particle rest mass, all Hawking-radiated particles may be approximated as ultra-relativistic. Thus the $kT_{BH}\gtrsim 2$ TeV instantaneous emission rate per fundamental particle species depends essentially only on the ratio $x = E/kT_{BH}$, as in Fig.~\ref{fig:primaryfluxes}: the dependence of the rate on $E$ is the same for different $\tau$ values except for a translation proportional to $T_{BH}$ (see Fig.~\ref{fig:quarkflux}) and the SEM Hawking radiation rate per degree of freedom has a scale invariance with respect to $x$.

This scale invariance leads to useful power law approximations.
For example, the direct Hawking emission rate for Dirac particles
with $x \ll 1$ is proportional to $x^2$ (recall Fig.~\ref{fig:primaryfluxes}) and the direct Hawking emission rate for vector $s=1$ particles with $x \ll 1$ is proportional to $x^{3}$ \cite{Page1976a}. Other power laws appear in the emission rate
for the final state photons that are created in the decays of the directly Hawking-radiated quarks and gluons.


\subsection{QCD Fragmentation}

According to the SEM, Eq.~\ref{eq:hawking}
applies to the direct Hawking radiation of the fundamental particles
of the Standard Model of high-energy physics: the leptons, quarks, and the gauge
bosons~\cite{Hawking1974, Hawking1975, Page1976a, Perry1977, MacGibbon1990}. As they stream away from the BH, these fundamental
particles will then evolve by Standard Model processes, into the
particles which are stable on astrophysical timescales. In particular,
quarks and gluons will undergo fragmentation and hadronization
into intermediate states which will eventually decay into photons,
neutrinos, electrons, positrons, protons and anti-protons. Because
the mean lifetime of a neutron at rest is
$\sim 10^3$ s, undecayed neutrons of high energy $E_{\rm N}$ should
also arrive from PBHs closer than $\sim (E_{\rm N}/100\ {\rm TeV})$ pc.
For application to PBH searches at VHE gamma-ray observatories,
we seek the total photon emission rate from the BH.
The photon production has several components\footnote{
Because we are investigating photon energies $\gtrsim 50$ GeV, we do not include the white inner
bremsstrahlung photon component generated by the Hawking radiation of charged fermions which is
dominated at these energies by the fragmentation photons~\cite{Page2008}.}:
(i) The ``direct photons'' produced by the direct Hawking radiation of photons:
this component peaks at a few times $T_{BH}$ and is most important at the highest photon energies at any given $T_{BH}$.
(ii) The ``fragmentation photons'' arising from the fragmentation and hadronization of the quarks and
gluons which are directly Hawking-radiated by the BH (in particular, quark and gluon fragmentation and
hadronization generates $\pi^{0}$'s which decay into 2 photons with a branching fraction of 98.8\%):
this component is the dominant source of photons at energies below $T_{BH}$.
(iii) The photons produced by the decays of other Hawking-radiated fundamental particles,
e.g., the tau lepton, $W$ and $Z$ gauge bosons, and Higgs boson;
this component is small compared to the component produced by the fragmentation of directly
Hawking-radiated quarks and gluons and is neglected here.
(We note that because the $W$, $Z$, and Higgs bosons decay predominantly via hadronic channels, their main
effect is to enhance the fragmentation photon component by at most $\sim 10\%$.)

In the SEM, the production rate of hadrons by the BH
is equal to the integrated convolution of the Hawking emission rates for the relevant fundamental particles $i$ (Eq.~\ref{eq:hawking})
with fragmentation functions $D_{h/i}(z)$ describing the fragmentation of species $i$ into hadron $h$, where
$z=E_{h}/E$ is the fraction of the initial particle's energy $E$ carried by the hadron; i.e.,
\begin{equation}\label{eq:convolution}
\frac{d^{2}N_{h}}{dE_{h} dt}
=\sum_{i} \int_{m_{h}c^2}^{\infty} \int_{0}^{1}
\ \frac{d^{2}N_{i}}{dE dt}
\ D_{h/i}(z) \; \delta(E_{h}-zE) \; dz \; dE.
\end{equation}
Here the summation is over all contributing fundamental species $i$ and
$D_{h/i}(z)dz$ is the number of hadrons $h$ with energy fraction in the
range from $z$ to $z+dz$ produced by the fragmentation of fundamental particle $i$.

For the current study, we wish to describe the photon burst generated in the final moments of the BH's
evaporation lifetime and the resulting light curve and energy spectrum seen by the detector. Fragmentation
functions $D_{b/a}(z)$ have been measured in high-energy physics experiments, such as $e^{+} e^{-}$ annihilation, for a variety of initial partons ($a$)
and final fragments ($b$)~\cite{deFlorian2007,Albino2008}.
However, a complete set of fragmentation functions is not available.
We turn therefore to a simplified fragmentation model. This model, which has appeared in the literature previously to estimate the photons derived from the
fragmentation of partons~\cite{Hill1983,Hill1987,Halzen1991},
is expected to provide a realistic representation of the photon spectrum for our purpose
and has been used in the analyses of PBH searches by several gamma-ray
observatories~\cite{Alexandreas1993, Linton2006}.
Alternatively, Eq.~\ref{eq:convolution} can
be evaluated using a Monte Carlo simulation which incorporates a parton showering program
such as Pythia~\cite{Pythiya2008} or Herwig~\cite{Herwig2013} extended to generate decays
into the astrophysically stable species, including photons. This approach has also previously
been used in PBH flux calculations~\cite{MacGibbon1990}
and is necessary if the goal is to obtain full spectral details about the instantaneous
flux of final-state particle species. We note that in both approaches, the $kT_{BH}\gtrsim 2$ TeV BH burst calculation requires extrapolation of the fragmentation functions or event generator codes to higher energies than have been validated in accelerator experiments.

\section{Photons from a BH Burst}\label{sec:burstphotons}

\subsection{The Pion Fragmentation Model} \label{subsec:pionfrag}

For photon production, the most important decay from the fragmentation of the initial quark or gluon is $\pi^{0} \rightarrow 2\gamma$.
In the pion fragmentation model, we proceed assuming that the QCD fragmentation of quarks and gluons may be approximated entirely by the production of pions. Two questions must be
addressed by the model: what is the pion spectrum generated by the partons and what is the
photon spectrum generated by the pion decays?

\begin{figure}
\begin{center}
\includegraphics[width=0.75\textwidth]{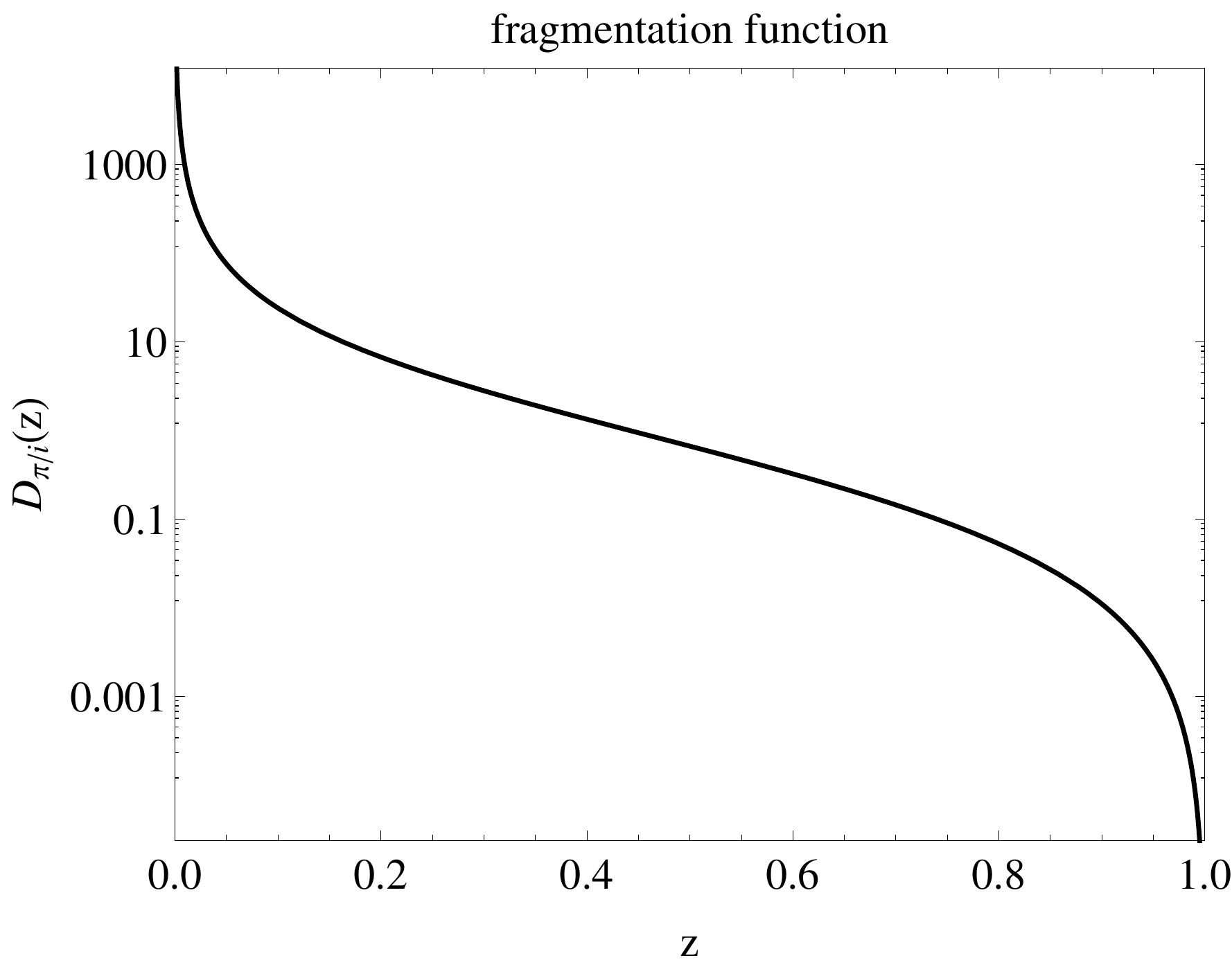}
\caption{The heuristic fragmentation function $D_{\pi /i}(z)$
as a function of the pion energy fraction, $z$.
\label{fig:fragmentation}}
\end{center}
\end{figure}

To answer the first question, we utilize a heuristic fragmentation function
\begin{equation}\label{eq:hff}
D_{\pi /i}(z)=\frac{15}{16}\ z^{-3/2}\ (1-z)^{2}
\end{equation}
where $z \equiv E_{\pi}/E$ is the energy fraction carried by a pion
generated by a parton of energy $E$~\cite{Halzen1991,Hill1983,Hill1987}. This function is normalized such that $\int_{0}^{1} z D_{\pi /i}(z) dz = 1$; i.e., all of the initial parton energy is converted to go into pions.
Fig.~\ref{fig:fragmentation} shows the fragmentation function $D_{\pi /i}(z)$.
The results in this paper are based on assuming the pion
fragmentation function $D_{\pi /i}(z)$ of Eq.~\ref{eq:hff} for all initial Hawking-radiated quarks and gluons. We note, though, that the function Eq.~\ref{eq:hff} implies an average energy for the final state photons and a multiplicity (number) of final state photons per initial parton which match the $T_{BH}^{1/2}$ scaling of the photon average energy and multiplicity found using a HERWIG-based Monte Carlo simulation to generate fragmentation and hadronization of the Hawking-radiated particles from $1\ \rm{GeV} \leq T_{BH} \leq 100\ \rm{GeV}$ black holes~\cite{MacGibbon1990}. We discuss the accuracy of this heuristic model further in Section~\ref{dis:pion_frag}.\footnote{
The function $D_{\pi /i}(z)$ resembles closely the result of a QCD calculation \cite{Hill1983}
which has been used for theoretical calculations in a previous PBH search \cite{Linton2006}.
Section~\ref{dis:pion_frag} has a discussion of the heuristic fragmentation function $D_{\pi /i}(z)$, compared to empirical fragmentation functions that have been extracted from collider data.} 

\begin{figure}
\begin{center}
\includegraphics[width=0.9\textwidth]{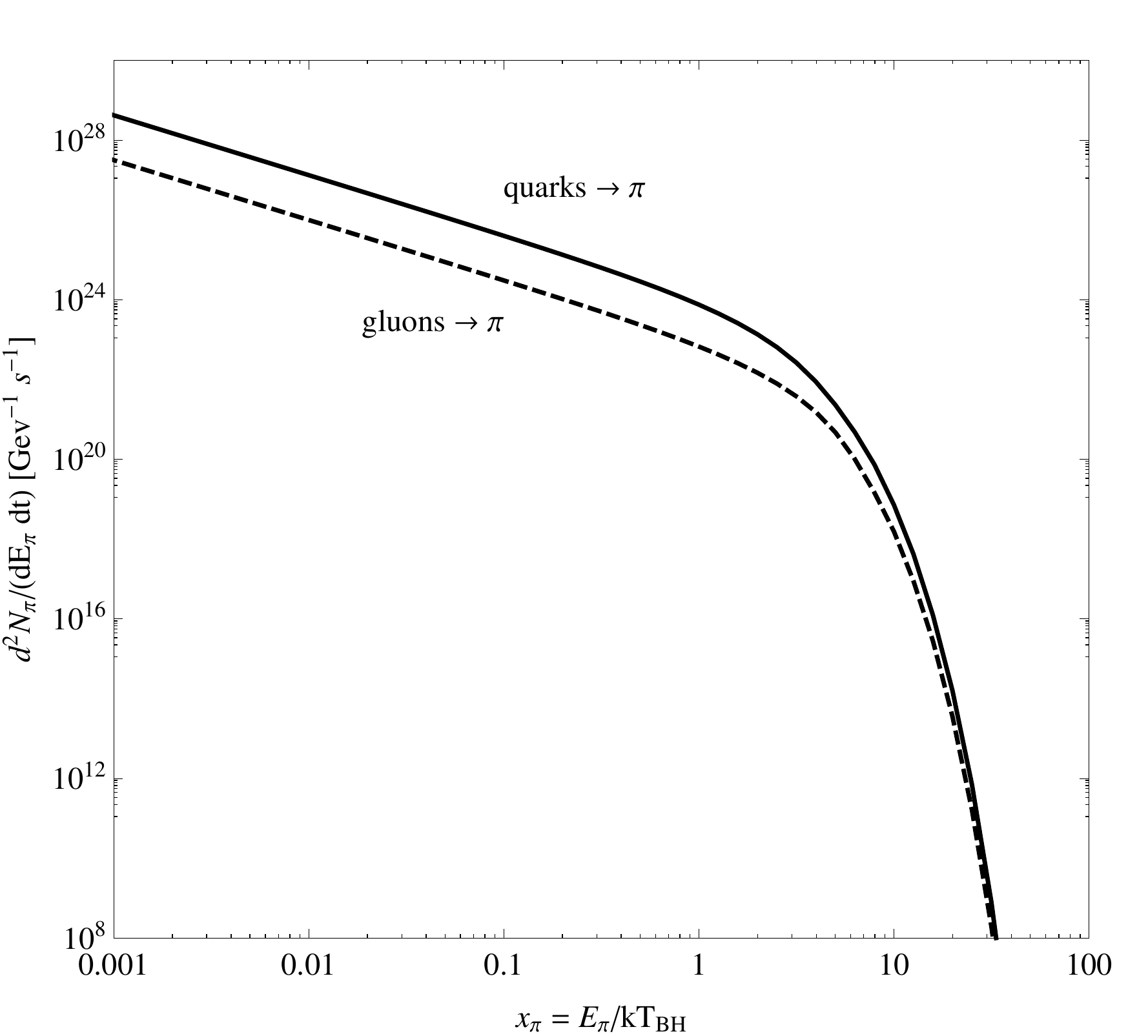}
\caption{The pion production rate $d^{2}N_{\pi}/(dE_{\pi} dt)$
calculated from Eq.~\ref{eq:d2Npi},
plotted as a function of $x_{\pi} = E_{\pi}/kT_{BH}$.
The number of degrees of freedom is $n_{\rm dof} = 72$ for quarks
and $n_{\rm dof}=16$ for gluons.
The quark contribution is dominant.
For $x_{\pi} < 1$ the particle rate obeys a power law:
$d^{2}N_{\pi} / (dE_{\pi} dt) \propto  x_{\pi}^{-3/2}$.
The power index of -3/2 matches the low-$z$ behaviour of
the assumed fragmentation function, which varies as $z^{-3/2}$.
\label{fig:pionflux}}
\end{center}
\end{figure}

The instantaneous pion production rate by the BH is then
\begin{equation}\label{eq:d2Npi}
\frac{d^{2}N_{\pi}}{dE_{\pi} dt} =
\sum_{i} \int_{m_{\pi}c^{2}}^{\infty} \int_{0}^{1}
 \frac{d^{2}N_{i}}{dE_i dt}\ D_{\pi /i}(z')\ \delta(E_{\pi}-z' E_i) dz' dE_i.
\end{equation}
Fig.~\ref{fig:pionflux} shows the instantaneous pion rate
as a function of $x_{\pi} \equiv E_{\pi}/kT_{BH}$.
At high pion energies, the pion rest mass is negligible and
this functon has a scaling form: it depends only on the
dimensionless ratio $x_{\pi}$.
(Similarly, we saw that the quark and gluon rates depend
only on $x = E/kT_{BH}$ when $T_{BH}$ is large
compared to the quark masses.)
Comparing Fig.~\ref{fig:primaryfluxes} and Fig.~\ref{fig:pionflux}
elucidates how the fragmentation of quarks and gluons at high energies,
say $E > 10-100\,{\rm TeV}$, yields a significant flux of pions at lower energies,
say $1\,{\rm GeV} < E_{\pi} < 100\,{\rm TeV}$; the $\pi^0$ decays then produce
photons in the detectable energy range of VHE gamma-ray observatories.

We assume that the pions are generated by the fragmentation and
hadronization of the 72 directly Hawking-radiated $s=1/2$ quark modes\footnote{Quarks come in 6 flavors, 3 colors, 2 spin states, and as particles or antiparticles for a total of 72 modes.}
and the 16 directly
Hawking-radiated $s=1$ gluon modes. The directly Hawking-radiated $W^{\pm}$ and $Z^{0}$ bosons also decay via hadronic jets about 70\% of the time.
Because $W^{\pm}$ and $Z^{0}$ are expected to each have only two polarization states
when $kT_{BH} \gtrsim 1\,{\rm TeV}$
(giving a total of 6 fundamental degrees of freedom)
and $\Psi_{s=1/2}\simeq 3\Psi_{s=1}$, however, the Hawking-radiated $W^{\pm}$ and $Z^{0}$ increase the instantaneous pion production rate of Fig.~\ref{fig:quarkflux} by only $\sim 3\%$.

Higgs modes also contribute to the pion flux to a small extent.
The dominant decay modes for the experimentally-confirmed $H^{0}$
resonance at $125$ GeV are $H^{0} \rightarrow b + \overline{b}$
and $H^{0} \rightarrow W^{+} + W^{-}$.
The dominant decay modes of any other Higgs states
(which have not yet been discovered) are also expected
to be $H \rightarrow q + \overline{q}$
and decays via $W^{\pm}$ and $Z^{0}$ bosons. Noting that $\Psi_{s=0}\simeq 2\Psi_{s=1/2}$, the directly Hawking-radiated Higgs boson states can
increase the instantaneous pion production rate of
Fig.~\ref{fig:pionflux} by at most $\sim 10\%$.

\subsection{Photon Flux from Pion Fragmentation}

We now obtain the photon flux from the $\pi^{0} \rightarrow 2\gamma$ decay of the pion distribution. Because the fragmentation function $D_{\pi /i}(z)$ includes
all three pion charge states $\pi^{+}, \pi^{-},$ and $\pi^{0}$ as equal components\footnote{
The charged pions do not contribute to the photon flux.
However, they do yield neutrinos.
The same heuristic model can be used to estimate the flux of neutrinos
from the BH.}, and each $\pi^{0}$ decays into two photons, we must multiply by 2/3 to get the $\gamma$ multiplicity.
In the $\pi^{0}$ rest frame,
the two photons have equal but opposite momenta and equal energies, $m_{\pi}c^2/2$.
In the reference frame of the gamma-ray observatory detector,
the energies of the two photons, $E_{\gamma}$, are unequal but complementary fractions
of the $\pi^{0}$ energy in the detector frame, $E_{\pi}$.
We assume that only one of the photons in each pair is detectable.\footnote{
The angle between the 2 photon trajectories in the detector frame will be very small because
of the large Lorentz boost.
However, if the BH is at a distance of order 1 parsec from the detector,
then only one of the photons from each $\pi^0$ decay will hit the detector.}

Let $\theta$ be the angle between the momentum of the observed photon
in the $\pi^0$ rest frame and the $\pi^0$ momentum in the detector frame.
In the detector frame, $E_{\gamma}=(E_{\pi} /2)(1+\beta\cos\theta)$
where the $\pi^{0}$ velocity $\beta = v/c\ \approx 1$
and $E_{\pi} = m_{\pi}/\sqrt{1-\beta^2}$. Because the angular distribution of the photons is isotropic in the $\pi^{0}$ rest frame,
the distribution of pion-produced photons in the detector frame is
\begin{equation}
\left(\frac{d^{2}N_{\gamma}}{dE_{\gamma} dt}\right)_{\rm frag.} =
\frac{2}{3} \int_{-1}^{1} \frac{2\pi d\cos{\theta}}{4\pi}
\int_{m_{\pi}}^{\infty}
\frac{d^{2}N_{\pi}}{dE_{\pi} dt}
\delta[E_{\gamma}-(E_{\pi}/2)(1+\beta \cos\theta)] dE_{\pi} .
\end{equation}
Evaluating the integral over $\cos\theta$, we have
\begin{equation}\label{eq:IntEpi}
\left(\frac{d^{2}N_{\gamma}}{dE_{\gamma} dt}\right)_{\rm frag.} =
\frac{2}{3} \int_{E_{\rm min}}^{\infty}
\frac{d^{2}N_{\pi}}{dE_{\pi}\,dt}
\frac{dE_{\pi}}{\sqrt{E_{\pi}^{2}-m_{\pi}^{2}}}.
\end{equation}
For $E_{\gamma} > m_{\pi}/4$, the minimum pion energy is $E_{\rm min} = E_{\gamma}+m_{\pi}^{2}/(4E_{\gamma})$.
Eq.~\ref{eq:IntEpi} implies that the photon energy
in the detector frame is uniformly distributed in the range
\begin{equation}
\frac{E_{\pi}(1-\beta)}{2} \leq E_{\gamma} \leq
\frac{E_{\pi}(1+\beta)}{2}.
\end{equation}
For high energy photons, we may approximate $m_{\pi} \approx 0$. In this case, the pion fragmentation function of Eq.~\ref{eq:hff} evolves into the photon distribution per initial parton
\begin{equation}\label{eq:gammaff}
D_{\gamma /i}(z_{\gamma})=\frac{15}{16} \bigg({\frac{16}{3}} + {\frac{2}{3}}z_{\gamma}^{-3/2} - 4z_{\gamma}^{-1/2} - 2z_{\gamma}^{1/2}\bigg)
\end{equation}
where $z_{\gamma} \equiv E_{\gamma}/E$ is the energy fraction carried by a photon
generated by a parton of energy $E$.

\begin{figure}
\begin{center}
\includegraphics[width=0.75\textwidth]{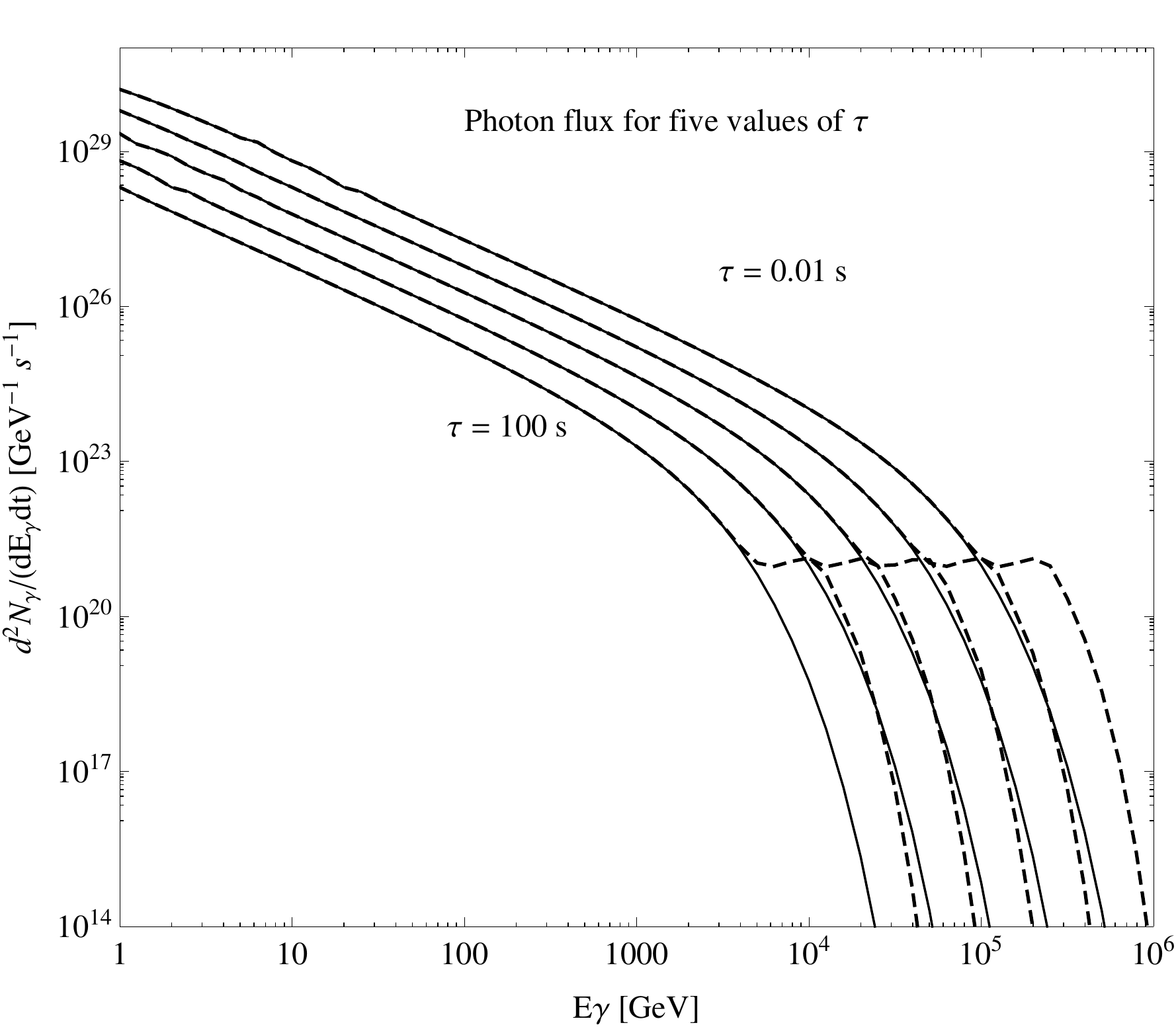}
\caption{The instantaneous photon emission rate
$d^{2}N_{\gamma}/(dE_{\gamma} dt)$ for five values of the remaining BH burst lifetime,
$\tau = 100$, $10$, $1$, $0.1$, and $0.01 \ {\rm s}$.
The dashed curves show the total (fragmentation + direct) gamma-ray emission rate;
the solid curves show the fragmentation only rate.
As $\tau$ decreases, the rate increases.
\label{fig:d2Ngamma}}
\end{center}
\end{figure}

Fig.~\ref{fig:d2Ngamma} shows the instantaneous
photon spectrum, including the directly Hawking-radiated photons, emitted by the black hole when the remaining BH evaporation lifetime is
$\tau = 100$, $10$, $1$, $0.1$ and $0.01 \ {\rm s}$.

%
%

\subsection{Parameterization of the Black Hole Photon Spectrum}\label{subsec:ParaSection}


To simplify subsequent calculations, we parameterize the function for the fragmentation component of the 
total photon emission rate from the BH as follows.
This parameterization is valid for $E_{\gamma}\gtrsim 1\ {\rm GeV}$.

By the scale invariance at short remaining time $\tau$,
$d^{2}N_{\gamma} / (dE_{\gamma} dt)$ depends on the ratio
\begin{equation}
x_{\gamma} = \frac{E_{\gamma}}{kT_{BH}(\tau)}
= 1.287 \times 10^{-4} \left(\frac{E_{\gamma}}{1\,{\rm GeV}}\right)
\left(\frac{\tau}{1\,{\rm s}}\right)^{1/3} .
\end{equation}
Fitting the curve shown in Fig.~\ref{fig:d2Ngamma}, we derive a reasonable parameterization of the fragmentation contribution to be
\begin{eqnarray}\label{eq:pbh_inst_para_1}
\left(\frac{d^{2}N_{\gamma}}{dE_{\gamma} dt}\right)_{\rm frag.}
&=&
A {x_{\gamma}}^{-3/2} \left[ 1 - \Theta_{S}(x_{\gamma}-0.3) \right]
\\
&+& B \exp(-x_{\gamma}) \left[ x_{\gamma}(x_{\gamma}+1) \right]^{-1}
\Theta_{S}(x_{\gamma}-0.3)
\end{eqnarray}
where
\begin{equation}
A = 6.339\times 10^{23} {\rm GeV}^{-1}{\rm s}^{-1},
~~~~~
B = 1.1367\times 10^{24} {\rm GeV}^{-1}{\rm s}^{-1}
\end{equation}
and
\begin{equation}
\Theta_{S}(u) = 0.5 \left( 1+\tanh(10u) \right).
\end{equation}
The accuracy of this parameterization is $\pm 15\%$ for $0.1\leq x_{\gamma} \leq 10$,
and $\pm 3\%$ for smaller and larger $x_{\gamma}$.
This is sufficient for most of our purposes. If greater accuracy is required, a table of the ratios of the approximate value to the exact value is used to correct the parameterized value.

We also derive, by curve-fitting, a parameterization of the directly Hawking-radiated photon component to be
\begin{equation}\label{eq:pbh_inst_para_2}
\left(
\frac{d^{2}N_{\gamma}}{dE_{\gamma} dt}\right)_{\rm direct}
=
\frac{(1.13 \times 10^{19}\ {\rm GeV}^{-1} {\rm s}^{-1}) (x_{\gamma})^6}
{\exp(x_{\gamma}) - 1}
F(x_{\gamma})
\end{equation}
where
\begin{equation}
F(x_{\gamma}) = 1.0 \ \ \ \ \rm{for}\ x_{\gamma} \leq 2
\end{equation}
and
\begin{equation}
\begin{split}
F(x_{\gamma})
=
\exp\left\{  \left[-0.0962-1.982\left(\ln{x_{\gamma}}-1.908\right) \right] \right.
\\
\left. \times\left[1+\tanh(20(\ln{x_{\gamma}}-1.908))\right] \right\}\ \rm{for} \ x_{\gamma} > 2.
\end{split}
\end{equation}


An alternative parameterization of the total photon emission rate by 
Linton et. al \cite{Linton2006}, which has often been used in PBH searches 
by high-energy observatories, is

\begin{equation}\label{eq:linton01}
\frac{d^2 N_{\gamma}}{dE_{\gamma} dt} = 6.24 \times 10^{23} \bigg[\frac{1}{8} \bigg(\frac{Q}{E_{\gamma}}\bigg)^{3/2}
-\frac{3}{4}\sqrt{\frac{Q}{E_{\gamma}}}-\frac{3}{8}\sqrt{\frac{E_{\gamma}}{Q}}+1\bigg] \, \rm{GeV^{-1} s^{-1}}, \,\,\,\,\text{for}\,\,\,\,\,E_{\gamma}<Q
\end{equation}

\begin{equation}\label{eq:linton02}
\frac{d^2 N_{\gamma}}{dE_{\gamma} dt} = 10^{21} \bigg(\frac{Q}{E_{\gamma}}\bigg)^4 \, \rm{GeV^{-1} s^{-1}}, \,\,\,\,\,\text{for}\,\,\,\,\,E_{\gamma} \geq Q
\end{equation}
\\
where $Q\simeq 4\times 10^4 (\tau /1\ \rm{s})^{-1/3}$ GeV is the energy of the peak quark flux averaged over the last $\tau$ seconds of the PBH's evaporation lifetime~\cite{Halzen1991}. The Linton et al. parameterization was derived by performing the convolution of Eq.~\ref{eq:d2Npi} with an approximation to the quark emission rate which replaces the energy dependence of Fig.~\ref{fig:quarkflux} with a delta function at $Q$.  Eq.~\ref{eq:linton01} is reasonably accurate for low energies but, for $E_{\gamma} > 0.1 Q$, differences of up to 30\% appear. More seriously, at $E_{\gamma}=Q$, the functional form of Eq.~\ref{eq:linton01} drops to
zero, which is not the actual behavior. Eq.~\ref{eq:linton02} matches the photon emission rate reasonably to within $20\%$ for $Q\leq E_{\gamma} < 2Q$ but strongly overestimates the exponential fall off in the emission rate at the highest energies.

\subsection{The Time-Integrated Photon Spectrum} \label{subsec:time-int}

In Section 3 we consider strategies for direct PBH burst searches at VHE gamma-ray observatories. One strategy is to utilize the photon time-integrated energy spectrum, i.e., the instantaneous photon emission rate integrated over a time interval
from an initial remaining evaporation lifetime $t=\tau$
to the completion of evaporation at $t = 0$, as a function of energy,
\begin{equation}\label{eq:TIS}
\left[\frac{dN_{\gamma}}{dE_{\gamma}}\right]_{\tau} =
 \int_{0}^{\tau} \frac{d^{2}N_{\gamma}}{dE_{\gamma} dt}\ dt.
\end{equation}

\begin{figure}
\begin{center}
\includegraphics[width=0.99\textwidth]{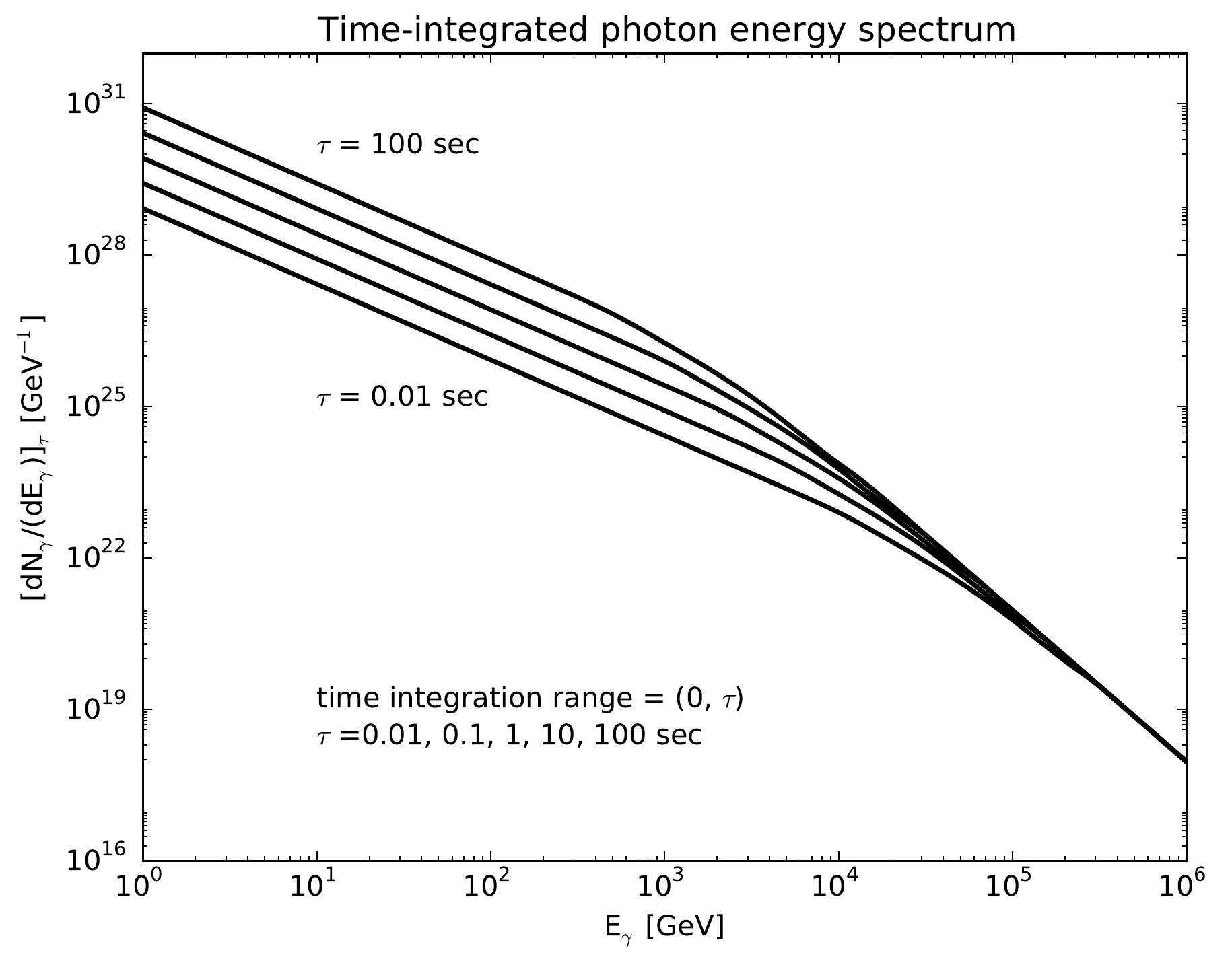}
\caption{The photon spectrum,
integrated over the final BH evaporation lifetime intervals
$\tau = 100, 10, 1, 0.1$ and $0.01 \ {\rm s}$.
\label{fig:photspect5DT}}
\end{center}
\end{figure}

Fig.~\ref{fig:photspect5DT} shows the photon time-integrated energy spectra for $\tau = 0.01$,
$0.1$, $1.0$, $10.0$, and $100.0\ \rm{s}$ using our parameterizations of Section \ref{subsec:ParaSection}.
In Fig.~\ref{fig:photspect5DT}, it can be seen that $\left[dN_{\gamma}/dE_{\gamma}\right]_{\tau}$
obeys different power laws above and below a transition energy
$E_{\rm tr}(\tau)$ which is of order $kT_{BH}(\tau) $: for $E_{\gamma} < E_{\rm tr}$,
$\left[dN_{\gamma}/dE_{\gamma}\right]_{\tau} \propto E_{\gamma}^{-1.5}$ and
for $E_{\gamma} > E_{\rm tr}$,
$\left[dN_{\gamma}/dE_{\gamma}\right]_{\tau} \propto E_{\gamma}^{-3.0}$.


To understand the origin of these slopes, we first change the variable of the integration in Eq.~\ref{eq:TIS} from $t$ to $x(t) = E/kT_{BH}(t)$ where 
$kT_{BH} (t) = E_0 (\tau_0 /t)^{1/3}$ is given by Eq.~\ref{eq:temptau} with $E_0 = 7800$ GeV and $\tau_0 = 1$ s. 
Thus $dt = 3\tau_0(E_0/E)^3 x^2 dx$ and
\begin{equation}\label{eq:pbh_int}
\bigg[\frac{dN_\gamma}{dE}\bigg]_{\tau} = 3\tau_0 \bigg(\frac{E_0}{E}\bigg)^3 \int_0^{x(\tau )}
\psi_{\gamma}(x_{\pi}')\ x'^2 dx'
\end{equation}
for a directly Hawking-radiated species. The integral in Eq.~\ref{eq:TIS} is dominated at high $E_{\gamma}$ by the directly 
Hawking-radiated photons and so Eq.~\ref{eq:pbh_int} implies 
$[dN_{\gamma}/dE_{\gamma}]_\tau \sim E_{\gamma}^{-3}$ for high $E_{\gamma}$; this result reflects the 
fact that approximately $\tau \propto T_{BH}^{-3}$. The integral in Eq.~\ref{eq:TIS} 
is dominated below $E_{\gamma} \sim kT_{BH}$ by the fragmentation function which must be convolved with Eq.~\ref{eq:pbh_int} and gives 
$[dN_{\gamma}/dE_{\gamma}]_\tau \sim E_{\gamma}^{-3/2}$ for $E_{\gamma}\lesssim kT_{BH}$;
this result reflects the fact that the fragmentation function Eq.~\ref{eq:gammaff}
behaves as $z_{\gamma}^{-3/2}$ for $z_{\gamma} \rightarrow 0$.

A parameterization for the time-integrated photon spectrum, derived by fitting the HERWIG-based Monte Carlo simulations of the photon flux from $1\ \rm{GeV} \leq T_{BH} \leq 100\ \rm{GeV}$ black holes of reference~\cite{MacGibbon1990}, was published by Bugaev et al.~\cite{Bugaev2007, Petkov2008},
\begin{equation} \label{eq:pbh_int_para}
\frac{dN_{\gamma}}{dE_{\gamma}} \approx 9 \times 10^{35}
\begin{cases}
\big(\frac{1\,{\rm GeV}}{T_{\tau}}\big)^{3/2}
\big(\frac{1\,{\rm GeV}}{E_{\gamma}}\big)^{3/2}
~~~~ {\rm GeV}^{-1}\ {\rm for} ~~ E_{\gamma} < kT_{\tau} \\
\big(\frac{1\,{\rm GeV}}{E{_{\gamma}}}\big)^{3}
~~~~ {\rm GeV}^{-1}\ {\rm for} ~~ E_{\gamma} \geq kT_{\tau}
\end{cases}
\end{equation}
for $E_\gamma \gtrsim$ 1 GeV. Here $T_{\tau}$ is the temperature of the black hole
at the beginning of the final burst time interval,
i.e., $T_{\tau} = T_{BH}(\tau)$ as given by Eq.~\ref{eq:temptau}. This approximation agrees well with our calculations of the time-integrated spectrum based on the pion fragmentation model, Eq.~\ref{eq:hff}, and
shown in Fig.~\ref{fig:photspect5DT}. Either could be used as input for
comparing the experimental sensitivities of different VHE gamma-ray observatories to a time-integrated PBH signal.


\subsection{BH Burst Light Curve}\label{subsec:pbh_lightcurve}

We now consider the time dependence of the BH burst,
i.e., the final ``chirp''. In Section~\ref{sec:pbh_search}, we explore whether knowledge of the time dependence can be used to enhance statistical significance in a PBH search.

\begin{figure}
\begin{center}
\includegraphics[width=0.99\textwidth]{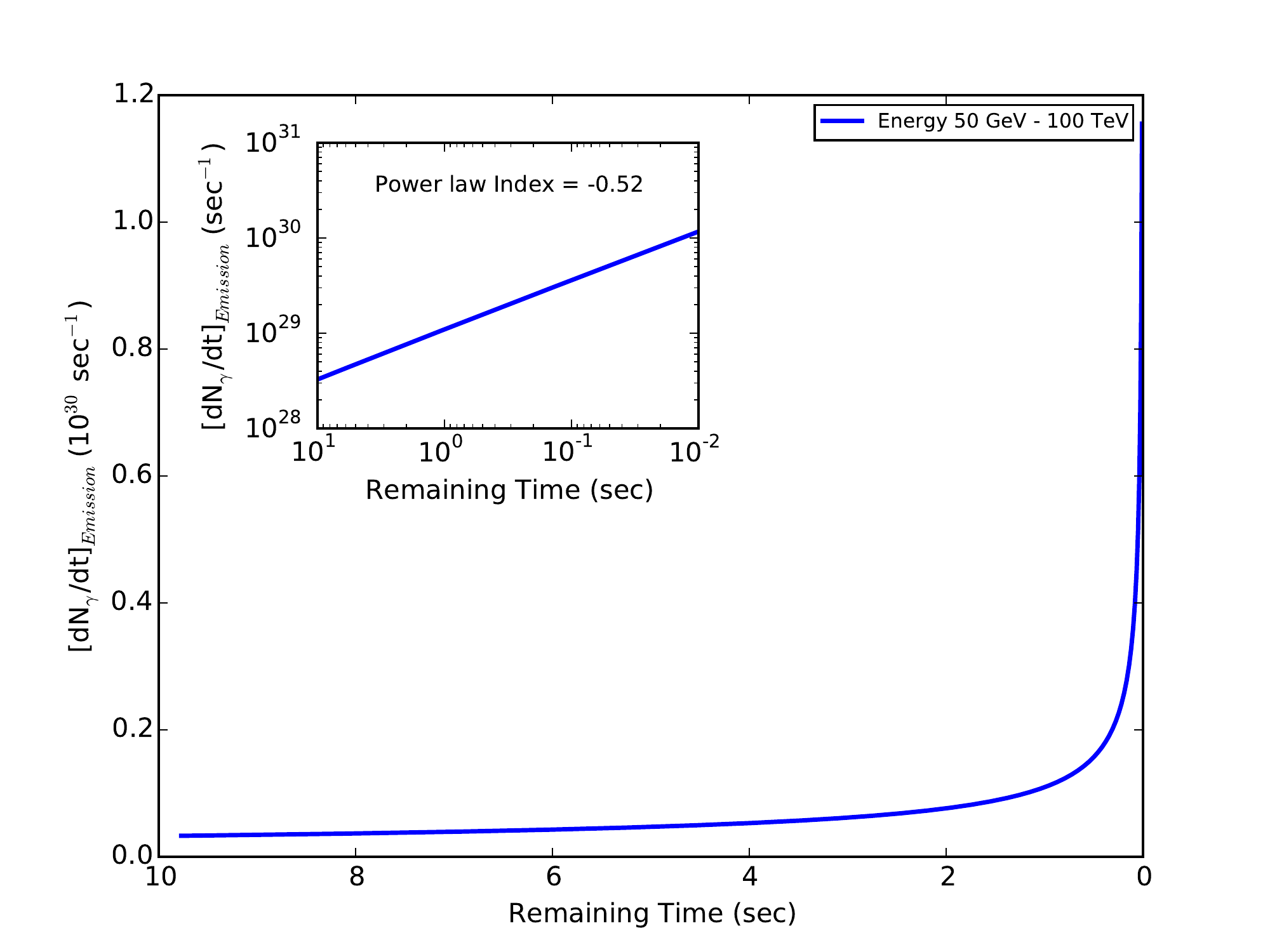}
\caption{The BH burst emission time profile, the emission rate integrated over energy, for the energy range 50 GeV -- 100 TeV.
As discussed in the text, this shape is well described by a power law with a
index of $\sim -0.5$.
\label{fig:pbh_lightcurve}}
\end{center}
\end{figure}

To find the time evolution of the BH burst, we integrate the differential emission rate
$d^{2}N_{\gamma}/(dE_{\gamma} dt)$ over photon energy $E_{\gamma}$
while retaining the time dependence. Hence the burst emission time profile (at the source) is
\begin{equation}\label{eq:LightCurve}
\bigg[\frac{dN_{\gamma}}{dt}\bigg]_{\text{Emission}}=\int_{E_{\rm min}}^{E_{\rm max}} \frac{d^{2}N_{\gamma}}{dE_{\gamma} dt} dE_{\gamma}.
\end{equation}
In general $E_{\rm min}$ and $E_{\rm max}$ are set by the energy range of the detector.
Fig.~\ref{fig:pbh_lightcurve} shows the BH burst emission time profile in
the energy range 50 GeV $\le E_{\gamma} \le$ 100 TeV.

It is interesting to relate the photon time profile to the total luminosity
function of the BH.  By basic thermodynamics, the luminosity
$dE/dt \propto T_{BH}^4 4 \pi R^2$ per emitted mode. For a Schwarzschild black hole, the radius $R \propto M_{BH}$ and $T_{BH} \propto 1/M_{BH}$, and so
$dE/dt \propto T_{BH}^2$ for the directly Hawking-radiated particles. Because the average energy of the directly Hawking-radiated particles is $\bar{E} \propto T_{BH}$, we expect for the directly Hawking-radiated particles that $dN/dt \sim {\bar{E}}^{-1} dE/dt \propto T_{BH}$. To estimate the $\tau$ dependence of $dN_\gamma/dt$, we note that the photon emission spectrum is dominated by the fragmentation component. The fragmentation function Eq.~\ref{eq:hff} implies a multiplicity (number of final states per initial particle) proportional to $T_{BH}^{1/2}$. Convolving the multiplicity dependence with the $dN/dt$ dependence per Hawking-radiated state leads to $dN_\gamma / dt \propto T_{BH}^{3/2}\propto \tau ^{-1/2}$, in agreement with the power law of approximately -0.5 found in Fig.~\ref{fig:pbh_lightcurve}. We also note that, because energy is conserved in the fragmentation and hadronization process, the total BH luminosity summed over all final state species is approximately $dE/dt \propto \tau^{-2/3}$.

\begin{figure}
\begin{center}
\includegraphics[width=0.9\textwidth]{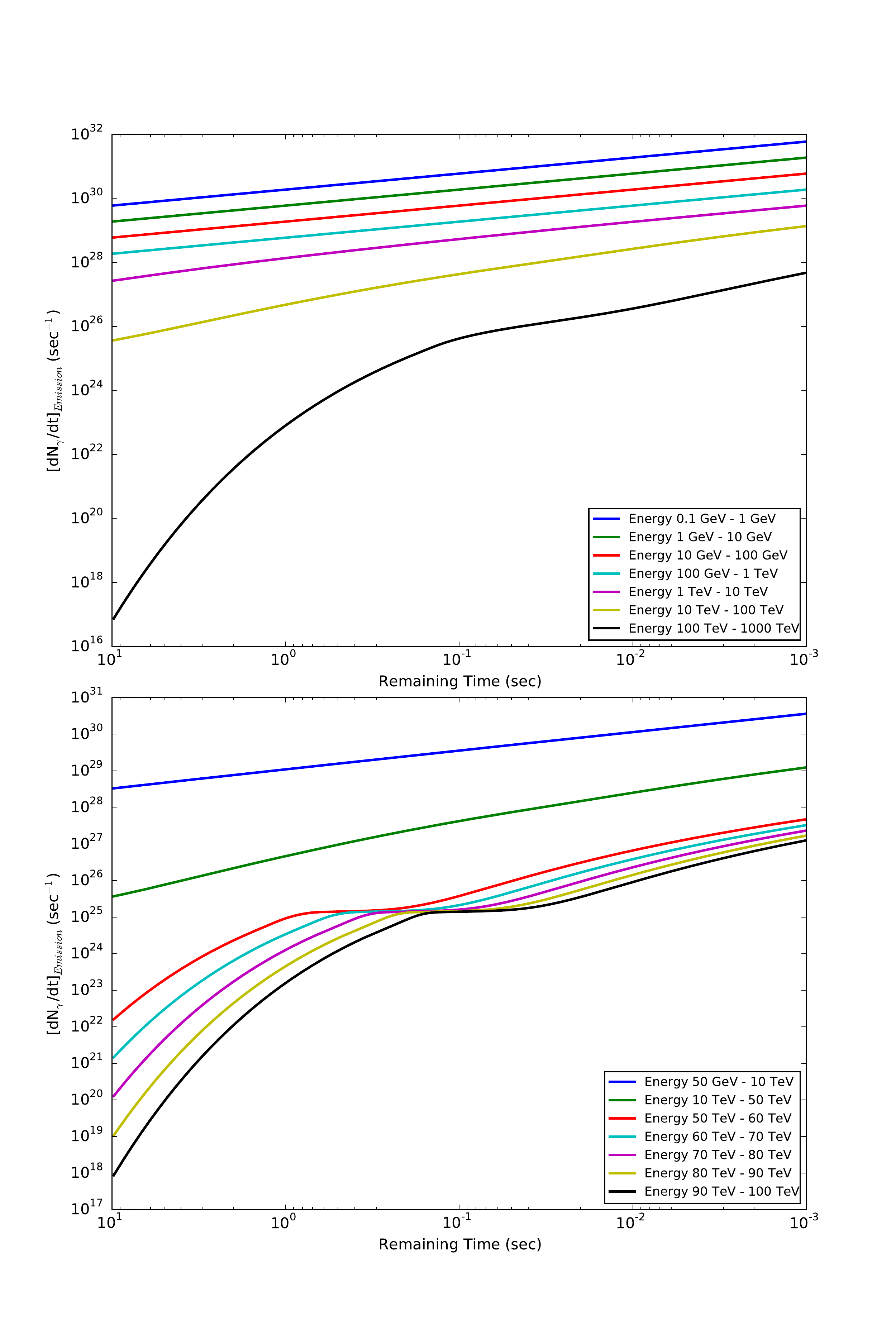}
\caption{Top panel shows the BH burst source emission time profile, the emission rate integrated over energy, for various integrated energy ranges. Here the shape of the emission time profile is the same for energy bands less than $\sim$ 10 TeV but is starting to become energy-dependent above $\sim$ 10 TeV. Bottom panel shows the  BH burst source emission time profile for a number of energy bands above 10 TeV. Here the energy dependence of the burst profile for higher energy bands can be clearly seen with an inflection region around $\tau\sim 0.1$ s.
\label{fig:pbh_multi_lightcurve}}
\end{center}
\end{figure}

The dependence of the BH burst emission time profile
on the energy range ($E_{\rm min}$, $E_{\rm max}$) is also relevant.
Fig.~\ref{fig:pbh_multi_lightcurve} shows $dN_{\gamma}/dt$ calculated using several ($E_{\rm min}$, $E_{\rm max}$) energy bands between 0.1 GeV and 1000 TeV.
We see that the low energy bands between $0.1$ GeV and $10$ TeV
have similar emission profiles. However, above energies of $\sim$ 10 TeV
the burst emission time profile is energy-dependent and has an inflection region
occurring $\sim 1$ s to 0.1 s before the end of the BH evaporation lifetime.
This energy dependence can be seen in the bottom panel of
Fig.~\ref{fig:pbh_multi_lightcurve}, where we have plotted several energy bands
above 10 TeV. The energy dependence of the burst emission time profile can be understood by
referring to Eq.~\ref{eq:pbh_inst_para_1} and Eq.~\ref{eq:pbh_inst_para_2} and Fig.~\ref{fig:d2Ngamma}. At low energies (below the inflection region), the photons
generated by the Hawking-radiated quarks and gluons dominate the flux;
at high energies (above the inflection point), the directly Hawking-radiated photons dominate the flux. In the inflection region, the two components are comparable.


Fig.~\ref{fig:pbh_lightcurve} and Fig.~\ref{fig:pbh_multi_lightcurve} display the emission time profiles, i.e., the source emission rate as a
function of remaining time. Let us now consider the detection time profile, i.e., the {\em light curve} of the PBH burst for a
specific VHE gamma-ray observatory. The detection time profile can be calculated from
\begin{equation}\label{eq:DetectorLightCurve}
\bigg[\frac{dN_{\gamma}}{dt}\bigg]_{\text{Detection}}=\frac{1}{4\pi r^{2}}
\int_{E_{\rm min}}^{E_{\rm max}} A(E_{\gamma})
\frac{d^{2}N_{\gamma}}{dE_{\gamma} dt} dE_{\gamma}
\end{equation}
where $A(E_{\gamma})$ is the effective area of the detector, as a function of photon energy $E$, and $r$ is the distance from the PBH to the detector.

\begin{figure} 
\begin{center}
\includegraphics[width=0.99\textwidth]{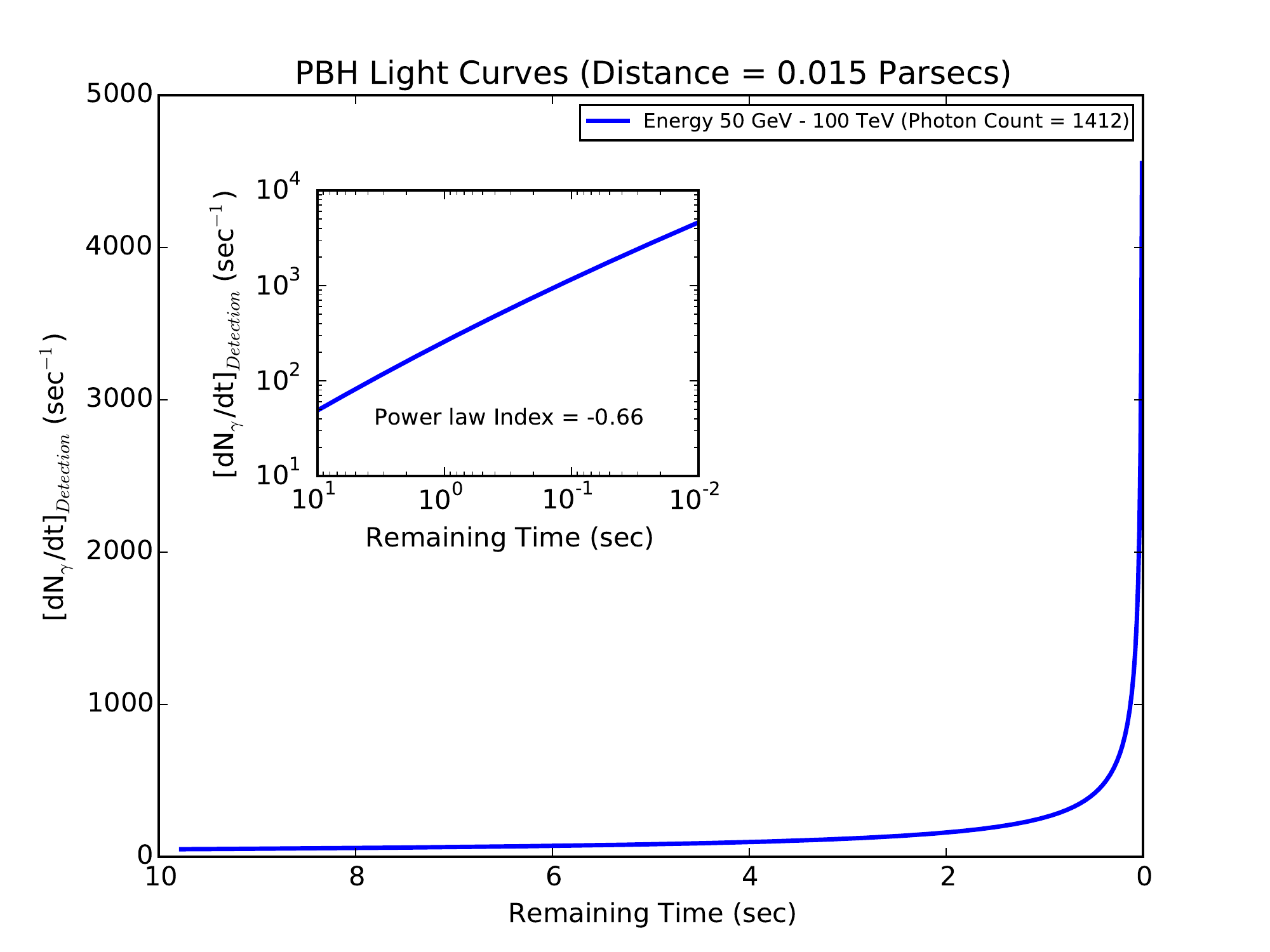}
\caption{Simulated PBH burst light curve observed by HAWC (at a distance of 0.015 parsecs)
obtained by convolving with the HAWC effective area published in Ref~\cite{Ukwatta_ICRC_PBH_2013}.
This shape is well described by a power law with a index of $\sim -0.7$.
\label{fig:pbh_lightcurve_hawc}}
\end{center}
\end{figure}

In this paper, we will use the HAWC observatory~\cite{UkwattaPBH2015a} as a
representative VHE gamma-ray observatory to investigate the PBH observational signatures.
Fig.~\ref{fig:pbh_lightcurve_hawc} shows the detection time profile for a PBH burst at a distance $r = 0.015$ parsecs, in the HAWC energy range (50 GeV --- 100 TeV) and for
the HAWC effective area published in reference~\cite{Ukwatta_ICRC_PBH_2013}. As discussed later, if the actual local PBH density is equal to the present limit on the local PBH density, HAWC might expect to have a $30\%$ chance of observing such a burst (at $0.015$ pc or closer) within its instantaneous field of view during 5 years of data taking.

An interesting feature of the BH signal is that the detection time profile (Fig.~\ref{fig:pbh_lightcurve_hawc}) rises more rapidly than the source emission time profile
(Fig.~\ref{fig:pbh_lightcurve}) as the remaining evaporation lifetime $\tau \rightarrow 0$.
This occurs because the effective area $A(E_{\gamma})$ is largest at high photon energies,
$E_{\gamma} > 10$ TeV, where both direct and fragmentation photon components are important but have different energy dependencies.

\begin{figure}
\begin{center}
\includegraphics[width=0.99\textwidth]{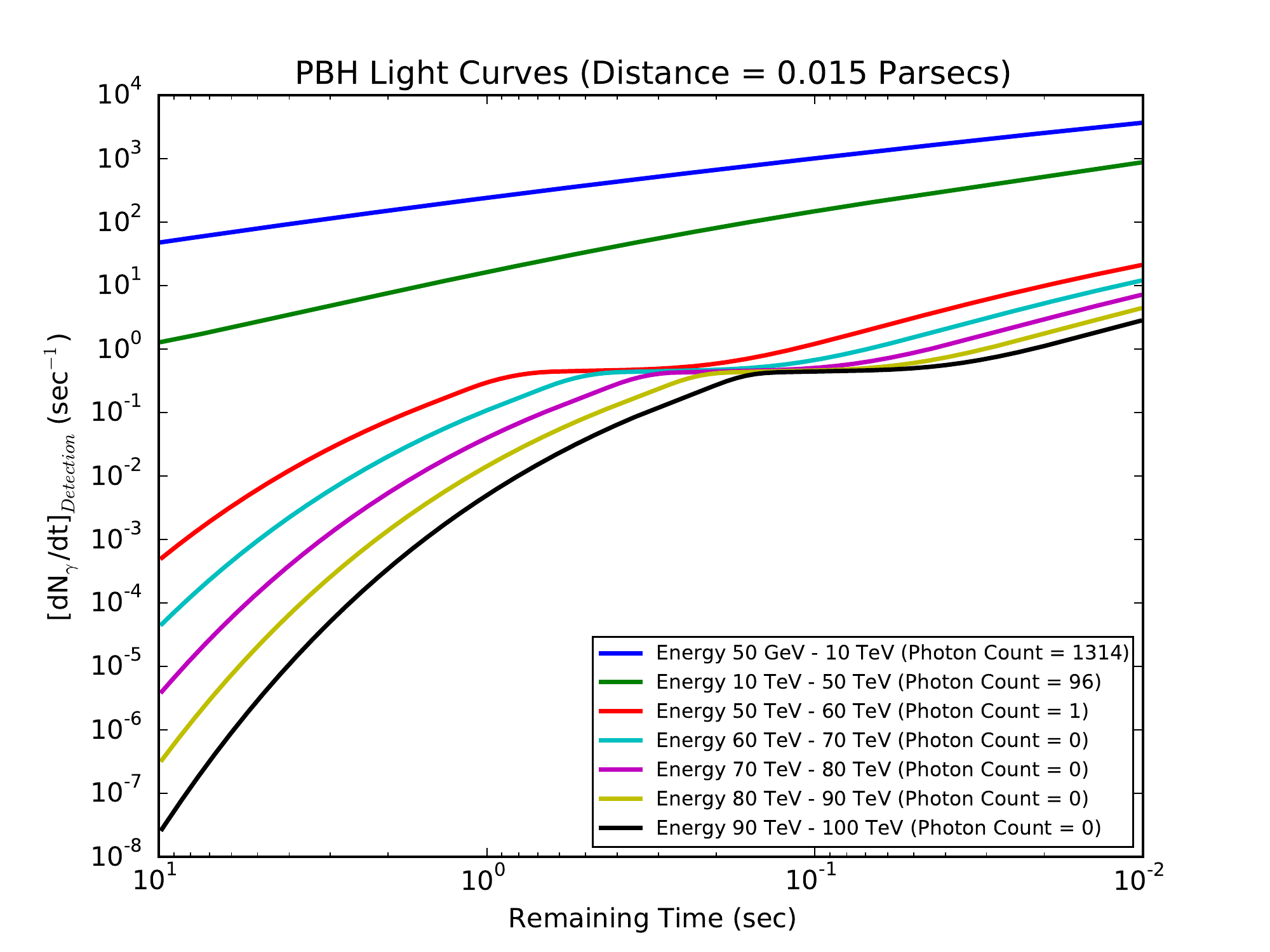}
\caption{Simulated detection time profiles of a PBH burst at a distance of 0.015 parsecs observed by 
HAWC in multiple energy bands. Photon numbers detected in each energy band are shown in the legend.
\label{fig:pbh_multi_lightcurve_high_hawc}}
\end{center}
\end{figure}

The HAWC detection time profiles for various photon energy ranges, analogous to the burst
emission time profiles shown in the bottom panel of Fig.~\ref{fig:pbh_multi_lightcurve},
are displayed in Fig.~\ref{fig:pbh_multi_lightcurve_high_hawc}.

\clearpage


\section{PBH Burst Searches and Upper Limits}\label{sec:pbh_search}

\subsection{PBH Burst Simple Search (SS) Method}\label{sec:simple_search}

The most straightforward way to search for a PBH burst (or any burst)
is to define search windows for the data in both time and space
(i.e., angular position on the sky), and then to inspect the search
windows for excess over the expected (sky and instrument) background;
the manner in which search intervals are defined, and
threshold levels set, varies. For a given detector, this Simple Search
(SS) method can be divided into two categories: the blind untriggered
SS and the externally triggered SS.
In a blind search, the time and location of the burst is 
not {\it a priori} known, and hence all temporal and spatial windows
need to be inspected for an excess over the background. This may
incur a large number of trials and correcting for them may reduce the
sensitivity of the search. An externally triggered search, on the other hand,
will look at a certain time and sky position for a burst once 
the burst has been detected in another
detector. Depending on how accurately the time and location 
of the externally triggered burst is known, a triggered search can incur 
typically one or a few trials, significantly less than a blind search.

\begin{figure}
\begin{center}
\includegraphics[width=0.9\textwidth]{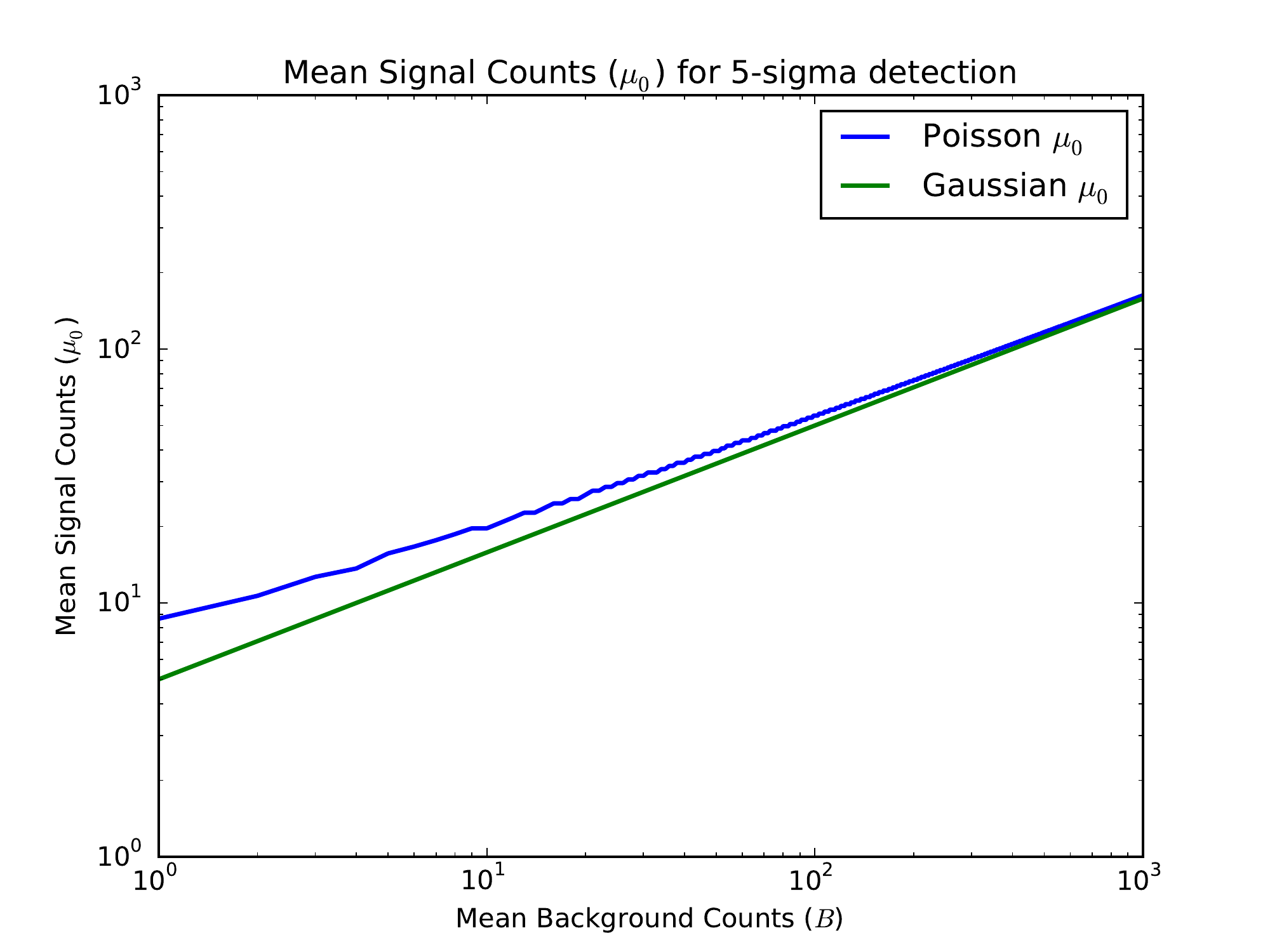}
\caption{Number of signal counts needed ($\mu_{\circ}$)  for a $5\sigma$ detection 50\% of
the time for a single trial as a function of the background counts using the PBH Simple Search (SS) method. The $\mu_{\circ}$
values are shown for a Poisson count distribution (upper blue line) and a Gaussian distribution (lower green line).
}\label{fig:signal_counts_needed_one_bin_plot}
\end{center}
\end{figure}

For both SS categories, we need to estimate the minimum number of expected
signal counts, $\mu_{\circ}$, required for a statistically significant burst detection to be probable.
The value $\mu_{\circ}$ may depend on the duration of the 
search window and its location in the detector's field. In order to
calculate $\mu_{\circ}$, fluctuations in both the background and the
signal need to be considered. Firstly, let $n$ be the number of
counts which has a probability of less than $p_0 \sim 2.87 \times
10^{-7}$ (corresponding to 5$\sigma$) of occurring under the 
background-only hypothesis, 
after correcting for $N_t$ trials. If
the detector counts follow a Poisson distribution and $B$ is the mean
number of background counts expected over the 
search window $\Delta t$, then $n$ is found from
\begin{equation} \label{stat1}
p_c = p_o /N_t = P(\geq n|B).
\end{equation}
where $p_c$ is the required p-value after correction for $N_t$
trials. The notation $P(\geq n|B)$ denotes the Poisson probability of getting
$n$ or more counts when the Poisson mean is $B$.
To estimate $\mu_{\circ}$, we also need to consider the fact that the
signal will fluctuate around some mean. Thus, we need to find the
mean value of the signal that together with the background
will give the desired $n$ counts in the detector X\% of the time.
The signal mean is then our
$\mu_{\circ}$ value. In typical searches, X\% is taken to be 50\%;
that is, $\mu_{\circ}$ is defined as the signal strength needed to
give a 5$\sigma$ detection 50\% of the time. Hence, for the SS
method, we can estimate $\mu_{\circ}$ from the equation
\begin{equation} \label{stat2}
P(\geq n | (B + \mu_{\circ})) = 0.5.
\end{equation}
Note that with these definitions, $\mu_{\circ} \approx n - B$ is 
reasonable approximation. Fig.~\ref{fig:signal_counts_needed_one_bin_plot} shows the mean signal counts,
$\mu_{\circ}$, for a 5$\sigma$ detection 50\% of the time for a single trial as a function of
background counts, based on the PBH SS method and a Poisson count distribution.
The $\mu_{\circ}$ values derived for a Gaussian distribution are also shown.

\subsection{Binned Maximum Likelihood Search (BMLS) Method}\label{sec:ml_search}

The Simple Search (SS) methods of Section~\ref{sec:simple_search}
use all photons detected in the search window irrespective of their
energy or time profile.
Thus, a SS does not make use of the time profile (the light curve
$dN_{\gamma}/ dt$) or the energy profile ($dN_{\gamma}/dE$) of the
PBH burst which we derived in Section~\ref{sec:pbh_theory}. Does
utilizing the time profile of the final seconds of the burst, 
we calculated in Section~\ref{subsec:pbh_lightcurve}, improve search
sensitivity? To address this question, we investigate the Binned
Maximum Likelihood Search (BMLS) method using a simple Monte Carlo simulation.\footnote{We also note that the energy profile of the
burst may improve the search sensitivity. For example, the background
of the detector may vary with the energy, and the detector
may be more sensitive in certain energy ranges. However, we defer
investigation of energy profile considerations to a separate paper.}


Consider a search window of duration $\Delta t$. Each search window
has an expected background mean ($B$), and possibly signal mean
($S$). The background counts are expected to be distributed over
$\Delta t$ with uniform probability while the PBH signal counts are expected to be distributed according to Fig.~\ref{fig:pbh_lightcurve_hawc}. In the BMLS method, we take
advantage of the PBH burst time profile by dividing each search
window into $k$ bins of time. If there is a PBH burst in a given
window then we expect the signal to be distributed in these $k$ 
according to Fig.~\ref{fig:pbh_lightcurve_hawc}.
For the purposes of comparison with the SS method, we will imagine
that for both the SS and BMLS the search window $\Delta t$ ends at
the expiration of the PBH burst, giving each search the best possible
alignment of the search window.
Thus, for a given search window, we can write the log of the likelihood ratio of the
signal-plus-background hypothesis to the background-only hypothesis as

\begin{equation} \label{eq:binned_llr}
\lambda = \sum_{i=1}^k \ln \frac{P(c_i|(s_i+b_i))}{P(c_i|b_i)}
\end{equation}
where $c_i$ is the observed number of counts in each bin, $s_i$ and $b_i$ are the
expected (mean) number of signal and background counts respectively in
each bin ($S = \sum_{i=1}^k s_i$ and $B= \sum_{i=1}^k b_i$), and $P(c|q)$ is the
Poisson probability of obtaining $c$ counts when $q$ counts are expected.

Proceeding further, we have
\begin{equation} 
\lambda = \sum_{i=1}^k \bigg( \ln \frac{(s_i+b_i)^{c_i} e^{-(s_i+b_i)}}{c_i!} - \ln \frac{b_i^{c_i} e^{-b_i}}{c_i!} \bigg) \nonumber 
\end{equation}
which simplifies to
\begin{equation} \label{eq:binned_llr4}
\lambda = \sum_{i=1}^k c_i \ln \bigg(\frac{s_i+b_i}{b_i}\bigg) - \sum_{i=1}^k s_i = \sum_{i=1}^k c_i \ln \bigg(1 + \frac{s_i}{b_i}\bigg) - S.
\end{equation}

\begin{figure}
\begin{center}
\includegraphics[width=0.99\textwidth]{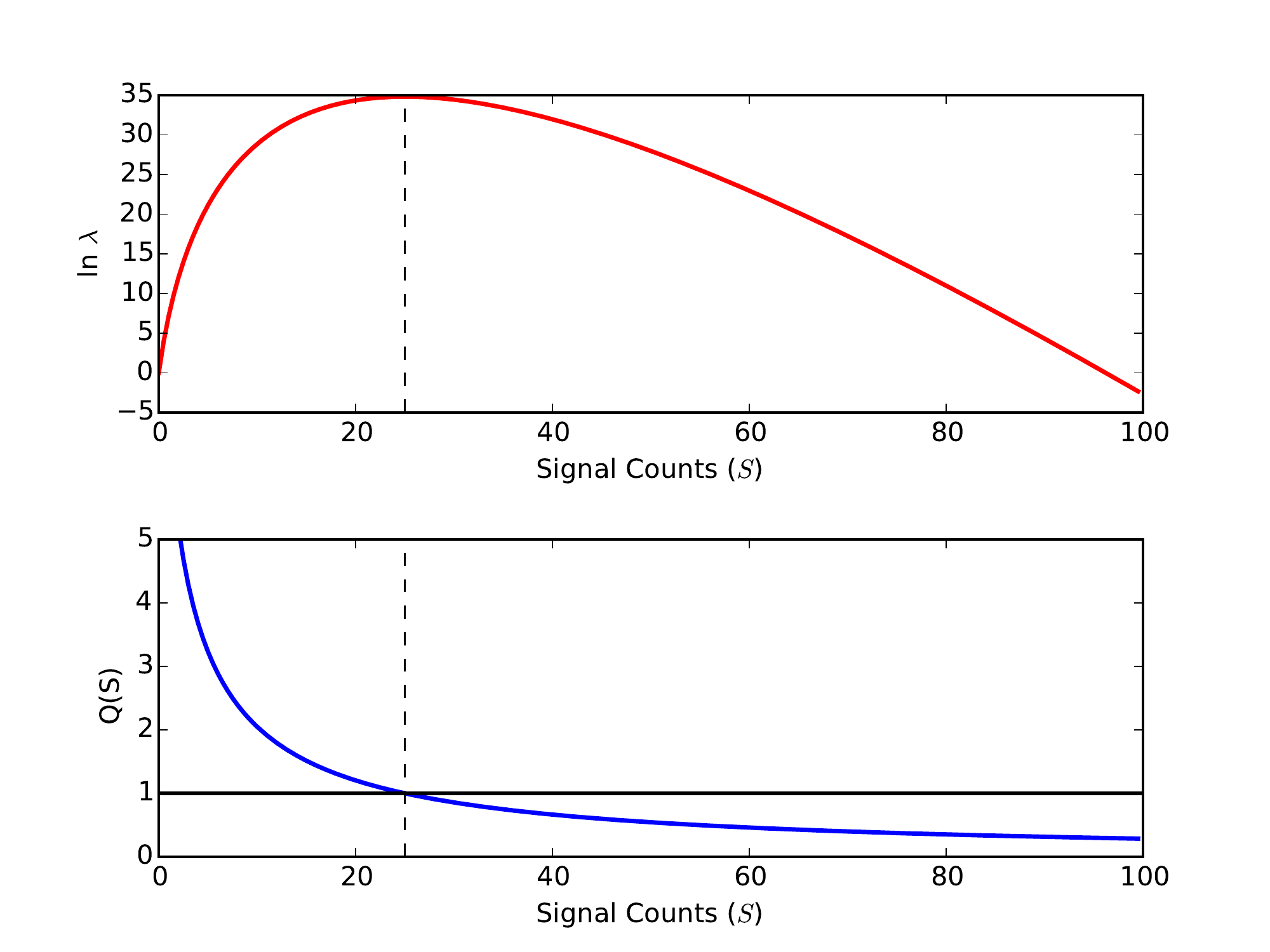}
\caption{Log likelihood ratio (upper panel) and $Q(S)$ (lower panel) as a
function of $S$ for a simulated Binned Maximum Likelihood Search (BMLS) bin with $\hat{S}$ = 25.
The log likelihood ratio is a maximum, i.e., $S = \hat{S}$, where the function $Q(S)$ = 1.
\label{fig:binned_ml_analysis}}
\end{center}
\end{figure}

If a non-zero signal occurs in a search window, then we estimate its strength as $\hat{S}$, the
value of $S$ which maximizes 
$\lambda$, as shown in the top panel of Fig.~\ref{fig:binned_ml_analysis}. (In
Fig.~\ref{fig:binned_ml_analysis}, $\lambda$ as a function of the expected signal $S$ is shown for
a single simulated randomized search window with $\hat{S}=25$.) The value of
$\lambda$ associated with $\hat{S}$ is denoted by $\hat{\lambda}$.
The maximum signal location can be found by partially differentiating
Eq.~\ref{eq:binned_llr4} with respect to $S$,

\begin{equation} \label{eq:binned_llr6}
\frac{\partial \lambda}{\partial S} = \sum_{i=1}^k \bigg(\frac{c_i}{s_i + b_i}\bigg) \frac{\partial s_i}{\partial S} - 1.
\end{equation}
Setting $s_i = S f_i$ where $f_i$ is the normalized binning of the PBH light curve shown in Fig.~\ref{fig:pbh_lightcurve_hawc}, we find
\begin{equation} \label{eq:binned_llr7}
\frac{\partial 
\lambda}{\partial S} = \sum_{i=1}^k \bigg(\frac{c_i f_i}{S f_i + b_i}\bigg) - 1 \equiv Q(S) - 1.
\end{equation}
For $
\lambda$ to be a maximum at $S = \hat{S}$, we require $\partial 
\lambda / \partial S = 0$ at $S = \hat{S}$ and so
\begin{equation} \label{eq:binned_llr9}
Q(\hat{S}) = \sum_{i=1}^k \bigg(\frac{c_i f_i}{\hat{S} f_i + b_i}\bigg) = 1.
\end{equation}
Because Eq.~\ref{eq:binned_llr9} is not analytically solvable, we
numerically evaluate $Q(S)$ and $Q(\hat{S})$, as shown in the bottom panel
of Fig.~\ref{fig:binned_ml_analysis}.
Noting $Q(S)$ is monotonically decreasing, we employed an efficient
binary search algorithm to find $\hat{S}$.

In order to find $\mu_0$, the mean number of expected signal counts necessary for a probable statistically significant
detection using the BMLS method, we need two levels of
simulation. First, we perform a background-only simulation to obtain the 
distribution of $\hat{\lambda}$ for background only, and determine $\lambda_c$, the value of $\hat{\lambda}$ corresponding to $p_c$, the p-value for $5\sigma$ significance, as for the SS method. 
We then perform a second set of simulations, varying the
expected signal mean until the BMLS finds $\hat{\lambda} > \lambda_c$ half the time, i.e., with a probability of 0.5. For the background-only
simulation, we use $\hat{\lambda}$ as a test statistic instead of the
signal strength $\hat{S}$ because, although $\hat{\lambda}$ and
$\hat{S}$ are correlated, the
log likelihood ratio $\hat{\lambda}$ more explicitly answers the question, ``is this search more signal-like than background-only-like?''.

For our BMLS, we chose $k$ time bins of equal duration.  In the
background-only simulation, we generate events in each bin by setting
$c_i \sim Pois (k^{-1} B)$, where $k$ is the number of bins, $B$ is the
expected background mean, and the notation $c_i \sim Pois(\beta)$
denotes that $c_i$ is a random number generated from a Poisson
distribution with mean $\beta$. For each iteration,
we find and record $\hat{\lambda}$. The procedure is repeated a large
number of times to find $\lambda_c$. This defines our criterion for a detection.

\begin{figure}
\begin{center}
\includegraphics[width=0.99\textwidth]{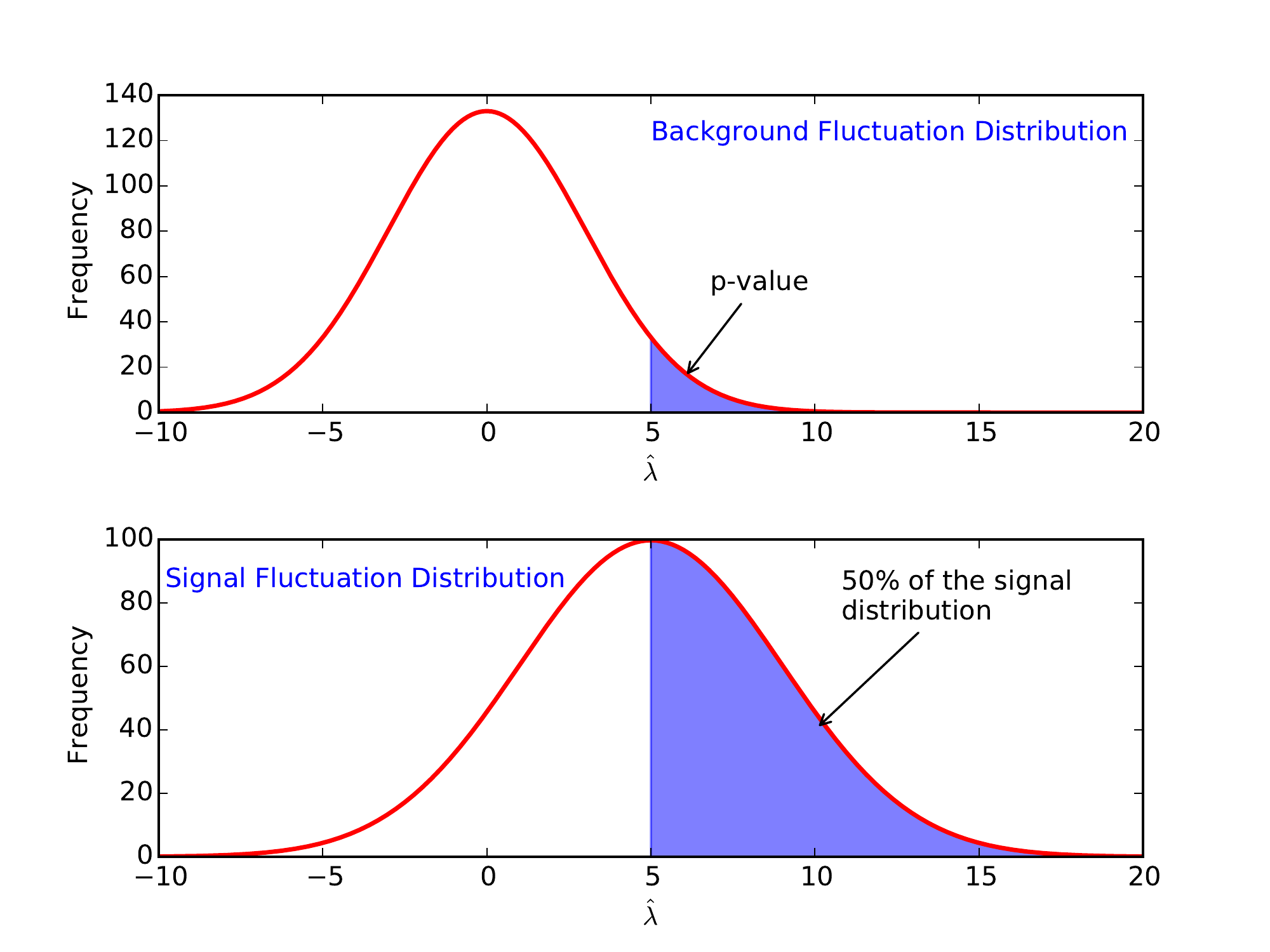}
\caption{
The number of mean signal counts ($\mu_0$) needed for a 5$\sigma$ detection
50\% of the time is found by first simulating many background-only searches.
From the resulting distribution of the maximum log likelihood $\hat{\lambda}$ values
(upper panel), we find the value required ($\lambda_c\ =\ 5$ in this case)
for the desired p-value. We then simulate many searches
containing both signal and background events. In the bottom panel
we show the distribution of the test statistic $\hat{\lambda}$ for the 
signal-plus-background case which produces 
$\hat{\lambda}\ >\  \lambda_c =\ 5$ with a probability of $50\%$.
A smaller value of the expected signal  $\mu_0$ would give a smaller fraction of searches
passing the $\hat{\lambda} > \lambda_c$ criterion.
\label{fig:signal_count_mc_plot}}
\end{center}
\end{figure}

For the signal-plus-background simulation, we run simulations with signal mean $S$
and vary $S$ until the search finds $\hat{\lambda} > \lambda_c$ $50 \%$ of the time.
The number of events in each bin is generated according to

\begin{equation} \label{eq:binned_llr11}
c_i^{sim} \sim Pois(S f_i + {\frac{B} {k}})
\end{equation}
The signal-plus-background simulation process is illustrated in Fig.~\ref{fig:signal_count_mc_plot}.
A good starting estimate for $S$ is the value
corresponding (on average) to $\lambda_c$, which itself can be estimated by
recording the $\hat{S}$ values during the background-only simulation.

\begin{figure}
\begin{center}
\includegraphics[width=0.99\textwidth]{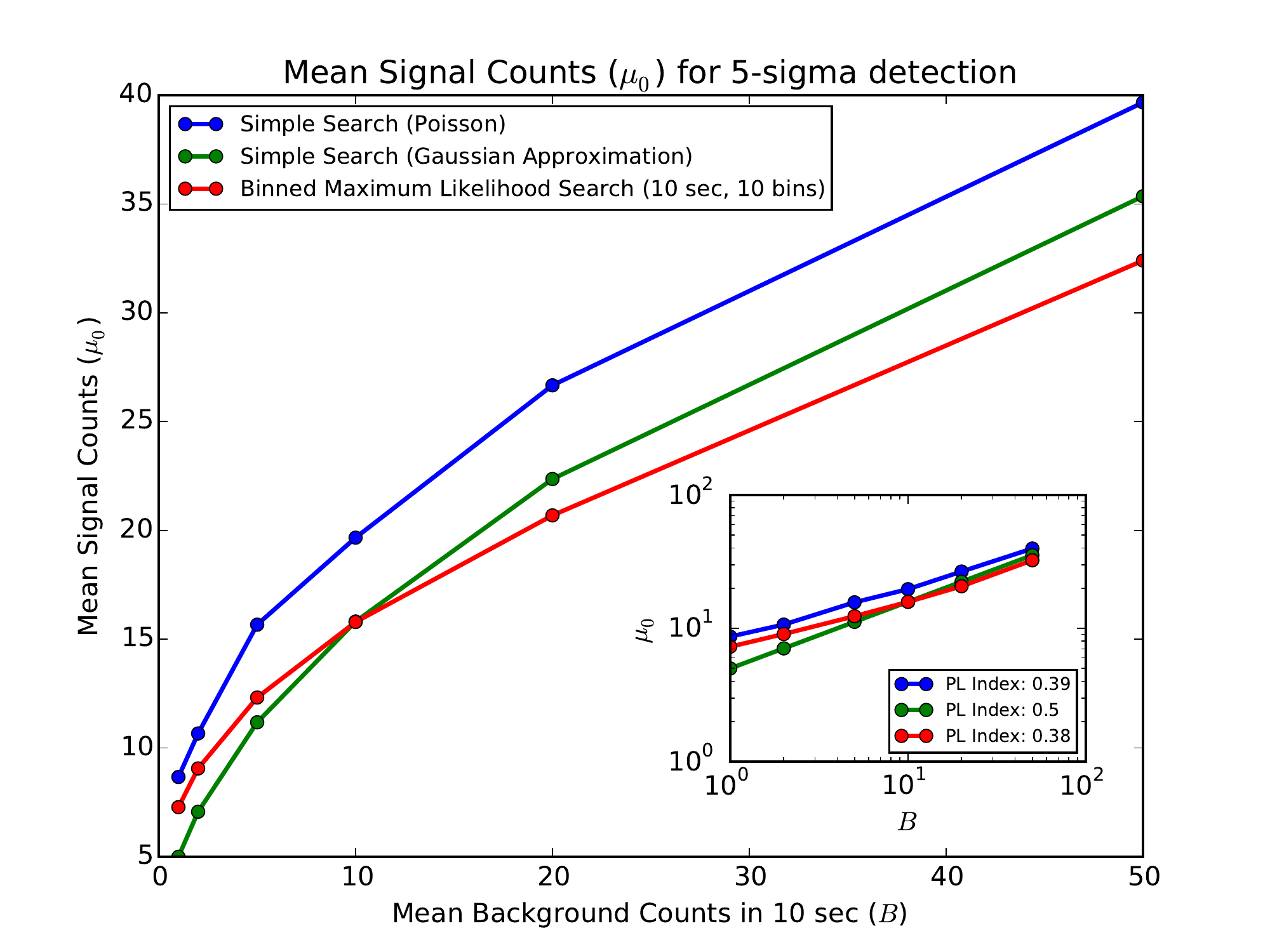}
\caption{Comparison of the two PBH burst search methods investigated in this paper:
the Simple Search (SS) method and the Binned Maximum Likelihood Search
(BMLS) method. The legend identifies the curves at large $B$ in top to
bottom order (blue, red, green). The insert shows the power law (PL)
indices of the dependence of $\mu_0$ on $B$.
Both search methods have similar power law dependencies.
\label{fig:signal_counts_needed_one_bin_binned_likelihood_plot}}
\end{center}
\end{figure}

The results of the BMLS simulation compared with the SS for PBH bursts
are shown in Fig.~\ref{fig:signal_counts_needed_one_bin_binned_likelihood_plot}.
For a value of $B$ corresponding roughly to the conditions of the HAWC 10 second expected
limit in reference~\cite{UkwattaPBH2015a}, the BMLS method would produce an upper limit
approximately a factor of 1.3 better than the SS method, using the
detector-related methods which we describe below in Section \ref{ul_estimate}~\cite{UkwattaPBH2015a}.

We have also investigated the Unbinned Maximum Likelihood Search method
using the complete unbinned time profile (``chirp'') of the PBH signal.
The unbinned search, however, results in little gain compared to a 
$k = 10$ bin search under the conditions of
the present study, namely moderate background events (less than 50) in the search window, and the simplifying assumption that we externally
know the burst time. Further studies are under way and will be reported in a separate paper.

\subsection{PBH Upper Limit Estimation}\label{ul_estimate}

In the case of a null detection (i.e., if no PBH bursts are observed), we can derive an upper
limit on the local PBH burst rate density, i.e., the number of PBH bursts per unit volume per unit time in the local solar
neighborhood. To calculate the upper limit, the PBH detectable
volume for a given detector is needed. The expected number of photons received by
a detector at or near Earth from a PBH burst during the last $\tau$ seconds of its evaporation lifetime at a non-cosmological distance $r$ and at detector angle $\theta$ is
\begin{equation} \label{eq:countsEq}
\mu(r, \theta, \tau) = \frac{(1-h)}{4 \pi r^2} \int_{E_1}^{E_2} \,\frac{dN_{\gamma} (\tau)}{dE_{\gamma}}\, A(E_{\gamma},
\theta)\,dE_{\gamma} .
\end{equation}
In this expression, $dN_{\gamma}(\tau)/dE_{\gamma}$ is the PBH photon emission energy spectrum
integrated from a remaining burst lifetime $\tau' = \tau$ to $\tau' = 0$.
We implicitly assume that the search window $\Delta t$ has been chosen to end at or near $\tau' = 0$. The function
$dN_{\gamma}(\tau)/dE_{\gamma}$ can be approximated using Eq.~\ref{eq:pbh_int_para}.
The energies $E_1$ and $E_2$ are the lower and upper bounds, respectively,
of the energy range of the detector, $h$ is the dead time fraction of the detector, and $A(E_{\gamma},\theta)$ is the
effective area of the detector as a function of $E_{\gamma}$ and $\theta$. The detector angle $\theta$ can be the zenith angle for
ground-based detectors or the bore sight angle\footnote{Bore sight angle is the angle between the pointing direction of the spacecraft and a source in its field-of-view.} for
space-based detectors. The function $A(E_{\gamma}, \theta)$ is typically obtained from a simulation of the detector.

In Sections~\ref{sec:simple_search} and ~\ref{sec:ml_search}, we estimated the
minimum number of expected signal counts needed for a detection, $\mu_{\circ}$.
By setting  $\mu_0$ equal to $\mu(r,\theta,\tau)$, the expected number
of counts from a PBH burst at a distance $r$ from Earth, we can solve Eq.~\ref{eq:countsEq} to find the maximum distance from which a given detector can detect
a PBH burst (with a detection probability of 50\%),
\begin{equation} \label{distanceEq}
r_{\rm max}(\theta, \tau) = \sqrt{ \frac{(1-f)}{4 \pi \mu_{\circ}} \int_{E_1}^{E_2} \,\frac{dN(\tau)}{dE_{\gamma}}\, A(E_{\gamma},
\theta)\,dE_{\gamma}}.
\end{equation}
The total PBH detectable volume of the detector is then
\begin{equation} \label{volueEq1}
V(\tau) = \sum_{\theta} V(\theta, \tau) = \frac{4}{3} \pi \sum_{\theta} r_{\rm max}^3(\theta, \tau) \frac{{\rm FOV_{\rm eff}}(\theta)}{4\pi},
\end{equation}
where FOV$_{\rm eff}(\theta)$ is the effective field-of-view associated with the detector angle
$\theta$ and the summation is over the bands of $\theta$ of the detector~\cite{UkwattaPBH2015a}.

If zero PBH bursts are observed, then the Y\% confidence level upper limit ($UL_{Y}$) on the rate density of PBH bursts, 
assuming that PBHs are uniformly distributed in the solar neighborhood, can be estimated as
\begin{equation}\label{ulX}
UL_{Y} = \frac{m}{V D}.
\end{equation}
Here $V$ is the effective PBH detectable volume, $D$ is the total search duration (typically in years) and $m$ is the upper limit at the
Y\% confidence level on the expected number of PBH events given that zero bursts are observed. Note that for Poisson fluctuations,
$P(0|m) = m^0 e^{-m}/0! = 1-Y$ implies that $m = \ln (1/(1-Y))$.
For $Y=99\%$, $m=\ln 100 \approx 4.6$ and hence the upper limit on the PBH
burst rate density in the case of null detection will be

\begin{equation}\label{ul99}
UL_{99} = \frac{4.6}{V D}.
\end{equation}

\begin{figure}
\begin{center}
\includegraphics[width=0.99\textwidth]{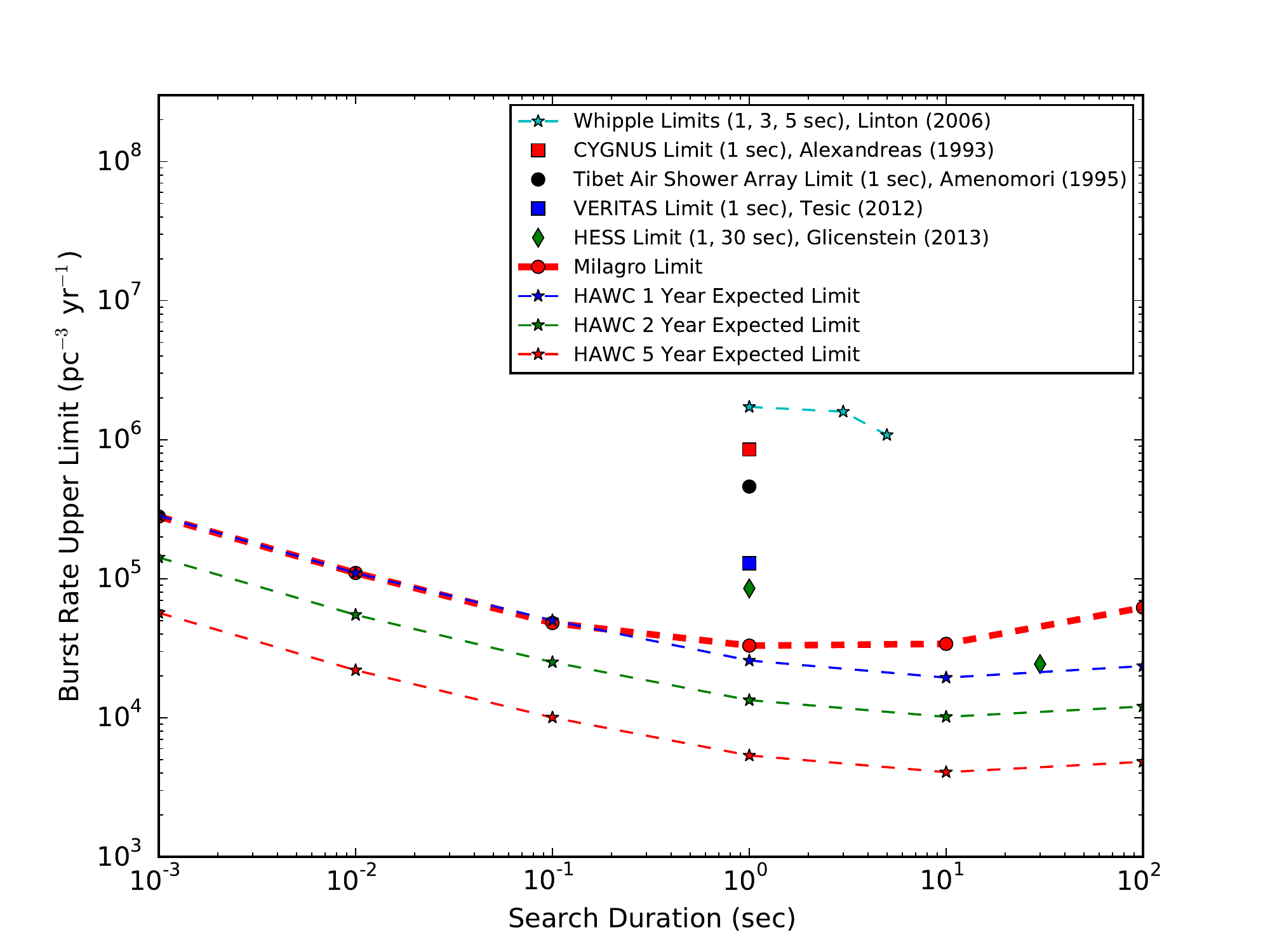}
\caption{Published PBH burst rate density $99\%$ CL upper limits and sensitivities for various 
experiments~\cite{UkwattaPBH2015a,Alexandreas1993,Amenomori1995,Linton2006,Tesic2012,Glicenstein2013,UkwattaMilagroPBH2013}. 
The upper limits and sensitivities shown are derived using the
Standard Emission Model description for the PBH emission spectra.
\label{fig:pbh_limits}}
\end{center}
\end{figure}

Fig.~\ref{fig:pbh_limits} shows published PBH burst rate density $99\%$ CL upper limits and sensitivities as a function of the search window, $\Delta t$, for various experiments. The PBH rate density limits calculated for Milagro and projected for HAWC are strictest around search window durations of 1 second and 10 second respectively. These optimum burst search intervals reflect the characteristics of the observatory and dependence of the background.

Let us understand the general features of Fig.~\ref{fig:pbh_limits}. From Equations \ref{distanceEq}--\ref{ul99}, we can see that $UL$, the upper
limit on the local PBH rate density, scales as
\begin{equation}\label{ul}
UL \sim (\mu_0(\Delta t)/N_{\gamma}(\tau)) ^{3/2}
\end{equation}
where $\mu_0$ is the sensitivity for a given search window $\Delta t$, and $N_{\gamma}(\tau)$
is the number of observable photons produced by the source over its remaining burst lifetime $\tau$.  Better sensitivity corresponds to smaller $\mu_0$, the number of signal photons required for detection of a signal, and a stricter
PBH limit.  For a source at a given distance (e.g., the one at the outer edge of the
volume considered), $N_{\gamma}(\tau)$ decreases when the search window is shorter, and
produces a weaker PBH limit. Shorter search windows incur less background events but also see fewer source photons. In this case, $\mu_0$ is
dominated by statistical fluctuations, in
particular those associated with the detector background rate, i.e.,
from Fig.~\ref{fig:signal_counts_needed_one_bin_binned_likelihood_plot}, $\mu_0 \propto B^{0.38} \propto (\Delta t)^{0.38}$; and $N_{\gamma}(\tau)$ is determined by the integral of the PBH time profile, slightly modified by the energy-dependent effective area of
the detector, i.e., from Fig.~\ref{fig:pbh_lightcurve_hawc}, $N_\gamma(\tau) \propto \int_0^\tau t^{-0.66}  \propto (\Delta t)^{0.34}$. These dependencies of $\mu_0$ and $N_\gamma(\tau)$ are both power laws and, despite their
very different physical origins, nearly cancel. Secondary effects such as the
larger number of trials incurred for shorter search windows ($N_t \propto 1/\Delta t$),
and the ability to optimize background rates for larger $\Delta t$ (for which the detection
efficiency is higher) give the $\Delta t$-dependence seen in Fig.~\ref{fig:pbh_limits}.

Thus, in summary, the shape of the PBH limit curves will be similar but will need to be
evaluated for each gamma-ray detector. The improvements suggested in this paper, of including the energy and time dependence of the PBH signal, will typically decrease the
required $\mu_0$. We anticipate that the PBH burst rate density
limit may particularly improve at longer search windows $\Delta t$ as a result.

\subsection{Differentiating the PBH Burst from Other GRBs}\label{sec:grb_vs_pbh}

Another important question regarding PBH burst detection is how to differentiate a PBH burst
from commonly detected GRBs of known cosmological origin. In particular, some
short GRBs of duration less than 2 seconds have light curves which are very similar to
the BH burst emission time profile shown in Fig.~\ref{fig:pbh_lightcurve}~\citep{Ukwatta_IPN_2015}.

Multi-wavelength observations are very important in differentiating PBH bursts from other known GRB source types.
Almost all GRBs have low-energy or VHE afterglows of recognizable shape which follow the main gamma-ray burst.
In the case of a PBH burst, no further signal is expected once the BH gamma-ray burst has expired,
with the possible exception of a signal generated by interaction of the charged emitted particles with the
ambient medium if the PBH is embedded in a sufficiently dense or turbulent ambient plasma
or magnetic field~\cite{MacGibbon2008}. The prompt burst time profile in various energy bands may also be used to distinguish a PBH origin.
Most GRBs show multi-peak structure with individual pulses 
exhibiting a Fast Rise Exponential Decay (FRED) shape. 
The light curve from a BH burst occurring in free space
is not expected to exhibit multi-peak structure at detector energies: for an isolated PBH
a single short peak described by Power-law Rise Fast Fall (PRFF) as shown in Fig.~\ref{fig:pbh_lightcurve} is expected. In addition, most GRB light curves show
hard-to-soft energy evolution while soft-to-hard energy evolution is expected with PBH burst light curves.

\begin{table}[htp]
\begin{center}
\begin{tabular}{|p{7cm}|| p{7cm}|}
\hline Gamma-ray Bursts (GRB) & PBH Bursts \\ \hline \hline
Detected at cosmological distances & Unlikely to be detected outside our
Galaxy  \\ \hline
Time duration can range from fraction of
second to few hours & Time duration is most likely less than few seconds \\ \hline
May have multi-peak time profiles & Single-peak time profile \\ \hline
Typically a single peak shows Fast Rise Exponential Decay (FRED) time profile
& Power-law Rise Fast Fall (PRFF) time profile expected\\ \hline
X-ray, optical, radio afterglows are expected & No multi-wavelength afterglow is expected \\ \hline
Most GRBs show hard-to-soft evolution & Soft-to-hard evolution is expected from
PBH bursts \\ \hline
Cosmic-rays are not expected to arrive from GRBs & Cosmic-ray bursts are expected from nearby PBH bursts \\ \hline
Gravitational wave signal is expected & No gravitational wave signal is expected  \\ \hline
Neutrino burst may be seen & Simultaneous neutrino burst may be seen from nearby PBH  \\ \hline
TeV radiation may be cut off either at the source or by the intergalactic medium & TeV signal is expected during the last seconds of the burst  \\ \hline
\end{tabular}
\caption{A summary of potential observational differences between standard cosmological GRBs and PBH bursts}
\label{table:pbh_vs_grb}
\end{center}
\end{table}

Moreover, with respect to TeV gamma-ray observations,
the extension of cosmological GRB spectra into TeV energies is uncertain
because of the attenuation of gamma-rays from distant GRBs by pair-production
off the intergalactic medium (IGM) and the possibility of a cutoff in the GRB source spectrum~\cite{Hauser2001,Albert2008}. In contrast, local
PBH bursts have a spectrum which extends well above 1 TeV during the latter parts of the BH
burst (for intervals as long as 100 s). Current instruments are sensitive to local PBH bursts ($<$ 1 parsec)~\cite{UkwattaPBH2015a}, where the ISM is not expected to
significantly attenuate TeV photons. Thus, detection in TeV observatories, together with
the other charactertistics expected for a BH burst, will lead to a potentially unique identification of the PBH signal.

In addition, GRBs due to PBH bursts are not expected to be accompanied by gravitational wave radiation. GRBs from other sources may be accompanied by gravitational waves (GW) and for short GRBs, a GW signature would confirm a compact star merger origin (which is the leading
model for short GRBs). Moreover, we expect the emission of a neutrino burst and cosmic-ray ($e^\pm$, $p$, $\bar{p}$ and possibly $n$) burst of similar time profiles to
accompany the gamma-ray radiation in the event of a BH burst. These neutrino and cosmic-ray
bursts should be emitted simultaneously by the BH with the gamma-ray burst. Thus far the reason that we have not detected neutrinos from standard GRBs may be due to their great distances. However, any PBH burst candidate that we detect with the current instruments should be very local, and so VHE neutrino and/or cosmic-ray telescopes may possibly detect the accompanying neutrino or cosmic-ray bursts from a PBH~\cite{Halzen1995, Smith2013, Keivani2015}. Table~\ref{table:pbh_vs_grb} summarizes the observational differences between standard cosmological GRBs and PBH bursts.


\section{BH Bursts with High-Energy Physics Beyond the Standard Model} \label{sec:BSM}

Our analysis in Section~\ref{sec:pbh_search} is based on the Standard Model of high-energy physics, in which the Hawking-radiated fundamental quanta are limited to those whose existence has been confirmed in high-energy experiments: the photon, neutrinos, charged leptons, quarks, gluons,
W and Z bosons, and the 125 GeV Higgs boson. There is strong evidence, however, that
the Standard Model is incomplete. For example, observations of neutrino oscillations
cannot be explained within the Standard Model and raise the question: are the neutrinos
Dirac fermions (with 4 degrees of freedom for each of the 3 neutrino flavors) or Majorana
fermions (with 2 degrees of freedom for each of the 3 neutrino flavors)? To date, only 6
neutrino degrees of freedom have been observed in detectors and so we assumed in Sections~\ref{sec:pbh_theory} - \ref{sec:pbh_search} that the neutrinos are Majorana fermions.

Other additional fundamental particle species may arise in Beyond the Standard Model (BSM) theories.
For example, supersymmetry (SUSY) would imply the existence of SUSY partners for all the known Standard
Model quanta: each $s=1/2$ fundamental Standard Model field would have an $s=0$ superpartner field and each
$s=1$ fundamental Standard Model field would have an $s=1/2$ superpartner field. Examples of other BSM
theories with additional degrees of freedom include extra dimension theories which imply massive
Kaluza-Klein excitations of known fields; shadow sector theories; and technicolor~\cite{Parsons2013,Chivukula2015}. In such models, the function
$\alpha(M_{BH})$ would increase at each new rest mass threshold, and thus the asymptotic rate of BH
evaporation would be faster than the SEM rate and the remaining BH lifetime would be shorter. The observable
spectra may be further modified,
depending on the degree to which the new particle species couple to ordinary matter and their decay
characteristics. Hence it is possible that BSM physics may modify our predictions for
the observation of the final stages of PBH evaporation. If new degrees of freedom
only manifest well above $100$ TeV, however, there will be little overall change to our analysis of
Sections~\ref{sec:pbh_theory} - \ref{sec:pbh_search}.

We note too that in this paper we are analyzing 4D black holes or black holes that
approximate 4D BHs. Higher-dimensional theories with $(n+4)$-dimensional gravity would have a lower Planck
mass. At very high $T_{BH}$ in such extra dimension theories, the equations relating black hole mass, temperature and remaining
lifetime and the emission spectra are significantly different to those of 4D BHs
once the BH Schwarzschild radius becomes smaller than that of the extra dimensions.
For recent reviews on accelerator limits on $(n+4)$-dimensional black holes see Refs.~\cite{CMS_ATLAS_BH2014, ATLAS_BH2014a, ATLAS_BH2014b}. 

In the following subsections, we address in greater detail the modifications to the 4D BH spectra that
would arise from Dirac neutrinos or SUSY.

\subsection{Dirac Neutrinos}

If neutrinos are Dirac fermions, the value of $\alpha$ must be modified.
We can estimate the effect at high $T_{BH}$ as follows. The total Hawking-radiated power determines the rate at which the
BH mass decreases. The function $\alpha(M_{BH})$, which is
defined by Eq.~\ref{eq:massloss}, accounts for the radiation of all relevant fundamental particle species
by an $M_{BH}$ black hole. For the detection of the final gamma-ray burst from a BH, we are interested in
remaining evaporation lifetimes in the range $\tau < 100$ s.
Including the $6$ extra degrees of freedom of Dirac neutrinos would increase
the asymptotic ($\tau < 100$ s) value of $\alpha(M_{BH})$, which we took in Sections~\ref{sec:pbh_theory} - \ref{sec:pbh_search}
to be $\alpha_{SM}$, by $12\%$ with little change to our analysis. 

If the extra Dirac neutrino degrees of freedom are light, $\alpha (M_{BH})$ would be increased at large $M_{BH}$ by a greater
percentage. In particular, if the rest masses of the extra Dirac neutrino modes are lighter than about $100$
MeV, the initial mass of a PBH whose lifetime equals the present
age of the universe would be larger than the SEM value of $5 \times 10^{11}$ kg by up to $15\%$.

\subsection{Black Hole Emission with Supersymmetry} \label{subsubsec:susy}

To examine the question, would the contributions from new BSM particles be discernable
at a VHE gamma-ray observatory if the observatory observes a PBH burst with a duration of $\tau\sim 100$ s, we now consider
a supersymmetric (SUSY) state to which HAWC might be sensitive. 
We focus on squarks (the $s=0$ superpartners of quarks)
because they represent a large number of degrees of freedom at a single threshold (due mainly to their
color degree of freedom), they decay into quarks which in turn decay into $\pi^0\rightarrow \gamma \gamma$ resulting in
observable TeV photons, and they are $s=0$ scalars with more intense Hawking flux and power than higher spin states.
The general statement, that the TeV photon emission rate increases at the rest mass threshold corresponding
to the new TeV particle mode, applies to whatever BSM theory is the origin of the new particle mode and hence
their consequences, while not identical, would resemble, or be weaker than, the squark radiation case.
The total Hawking flux and power will depend on the number of degrees of freedom and spin of the new
particle mode. For example, extra dimensional theories as a class may produce numerous new states, some of
which are colored. The effects of colored states would likely resemble those of squarks, but have a weaker
influence on the final photon spectra
unless the states were also $s=0$ scalars.  Non-colored states such as gauge particles or lepton-like particles
would be expected to have less effect on the TeV photon spectra
because of their fewer degrees of freedom, higher spin, and/or less frequent photon decays.

In SUSY models, such as the minimal supersymmetric standard model (MSSM),
there are many new fundamental superpartner fields.
Let us consider the case of a superpartner that is a squark of mass $m_{sq}$.
Once $kT_{BH}\gtrsim T_{sq} \equiv m_{sq}c^2/x_{p,s=0}$,
this squark species will appear in the Hawking radiation in significant numbers.
The squark will then decay into a quark and other particles, with
the quark then fragmenting and producing photons as described in Section~\ref{sec:pbh_theory}.

Let $\tau_{sq}$ be the remaining burst lifetime when the BH temperature
reaches $T_{sq}$. If $\tau_{sq}$ is much larger than 100 s---the time window
of the search---then a BH burst which includes squark emission
cannot be distinguished from an SEM BH burst because the distance
to the observed BH is undetermined: a BSM BH burst with squark radiation would produce approximately the same
signal in the detector as a closer SEM BH burst.

On the other hand, if $\tau_{sq} < 100$ s,
the observatory may witness the photon rate increasing
due to squark radiation as the remaining BH evaporation lifetime becomes less than $\tau_{sq}$,
provided the burst flux in the detector is large enough.

In many SUSY models, such as those tested at the LHC,
the rest masses of the superpartners are assumed to be of the order 500 GeV to 1 TeV, which correspond
to threshold times $\tau_{sq}$ much greater than $100$ s. (If the squark mass is of order 5 TeV, then 
$kT_{sq} = 1.9$ TeV and the threshold time $\tau_{sq}\lesssim 70$ s (from Eq.~\ref{eq:tautemp})
would occur during the time window of the PBH search.) The value of the remaining burst lifetime $\tau_{sq}$,
though, depends on $\alpha(M_{BH})$. If we assume that the evaporation process is dominated by SEM particles
until $T_{BH}$ reaches $T_{sq}$, then 
\begin{equation} \label{eq:presqmassloss}
\alpha(M_{BH}) = \alpha_{SM} \ \ 
{\rm ~~ for ~} 10^{8} ~{\rm kg}\gtrsim M_{BH} > M_{BH,sq}
\end{equation}
where $M_{BH,sq}$ is the BH mass threshold corresponding to $T_{sq}$.
(For a 5 TeV squark, $M_{BH,sq} = \hbar c^3 /(8 \pi G kT_{sq}) = 5.5 \times 10^{6} \, {\rm kg}$.)
For $M_{BH} < M_{BH,sq}$, $\alpha(M_{BH})$ increases by the contribution due to the squark emission, i.e.,
\begin{equation} \label{eq:sqmassloss}
\alpha(M_{BH}) = \alpha_{SM} + \alpha_{sq} \ \ 
{\rm ~~ for ~} M < M_{BH,sq} .
\end{equation}
Here the contribution from the squark degrees of freedom is
\begin{equation} \label{eq:sqalpha}
\alpha_{sq} = 1.7 \times 10^{17} ~~ {\rm kg}^{3}{\rm s}^{-1} 
\end{equation}
noting that the number of degrees of freedom for a squark with left-handed and right-handed  states of the same mass is 12 (particle, antiparticle, handedness, and 3 color modes) and $\Phi_{s=0}=6.89$ for each squark degree of freedom.

\begin{figure}
\begin{center}
\includegraphics[width=0.99\textwidth]{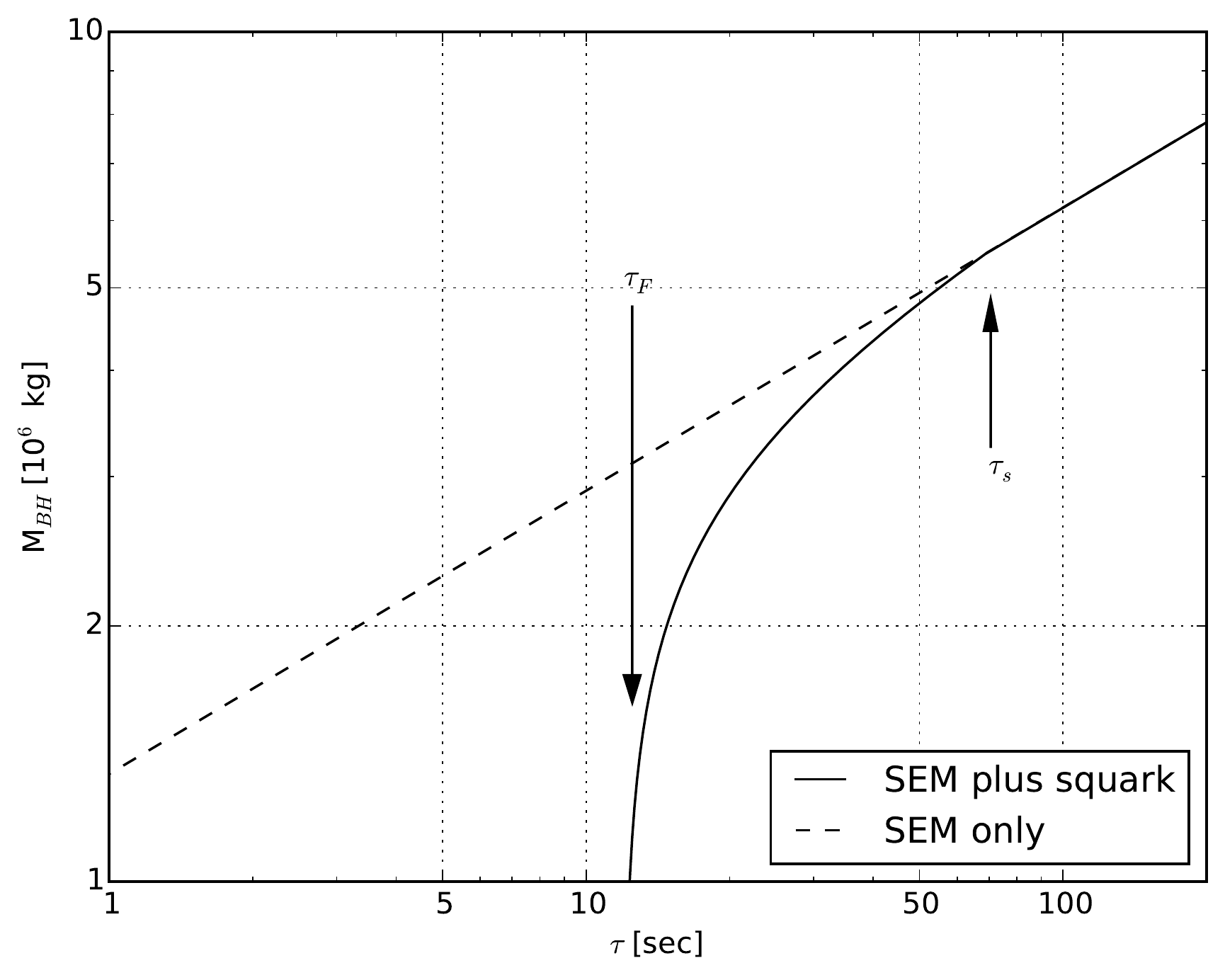}
\caption{Black hole mass of an SEM BH and a BSM (SUSY) BH as a function of the remaining lifetime of the SEM BH.
The time threshold for squark emission by the BSM BH is $\tau_{sq}= 70$ s;
the completion of the BSM BH burst occurs at $\tau_{F}= 12$ s.
Solid curve: the BSM black hole with squark radiation.
Dashed curve: the SEM black hole.
\label{fig:FIGY1}}
\end{center}
\end{figure}

Fig.~\ref{fig:FIGY1} shows the masses of an SEM BH and the BSM BH as a function of the remaining lifetime $\tau$ of the SEM black hole, assuming for alignment purposes that both BHs had the same mass when $\tau \gg \tau_{sq}$. In this example, the threshold time for emission of the squark is $\tau_{sq} = 70$ s and the BSM BH burst expires at $\tau_{F} = 12$ s. The value of $\alpha(M_{BH})$ is given by Eqs.~\ref{eq:presqmassloss} and~\ref{eq:sqmassloss}.

\begin{figure}
\begin{center}
\includegraphics[width=0.99\textwidth]{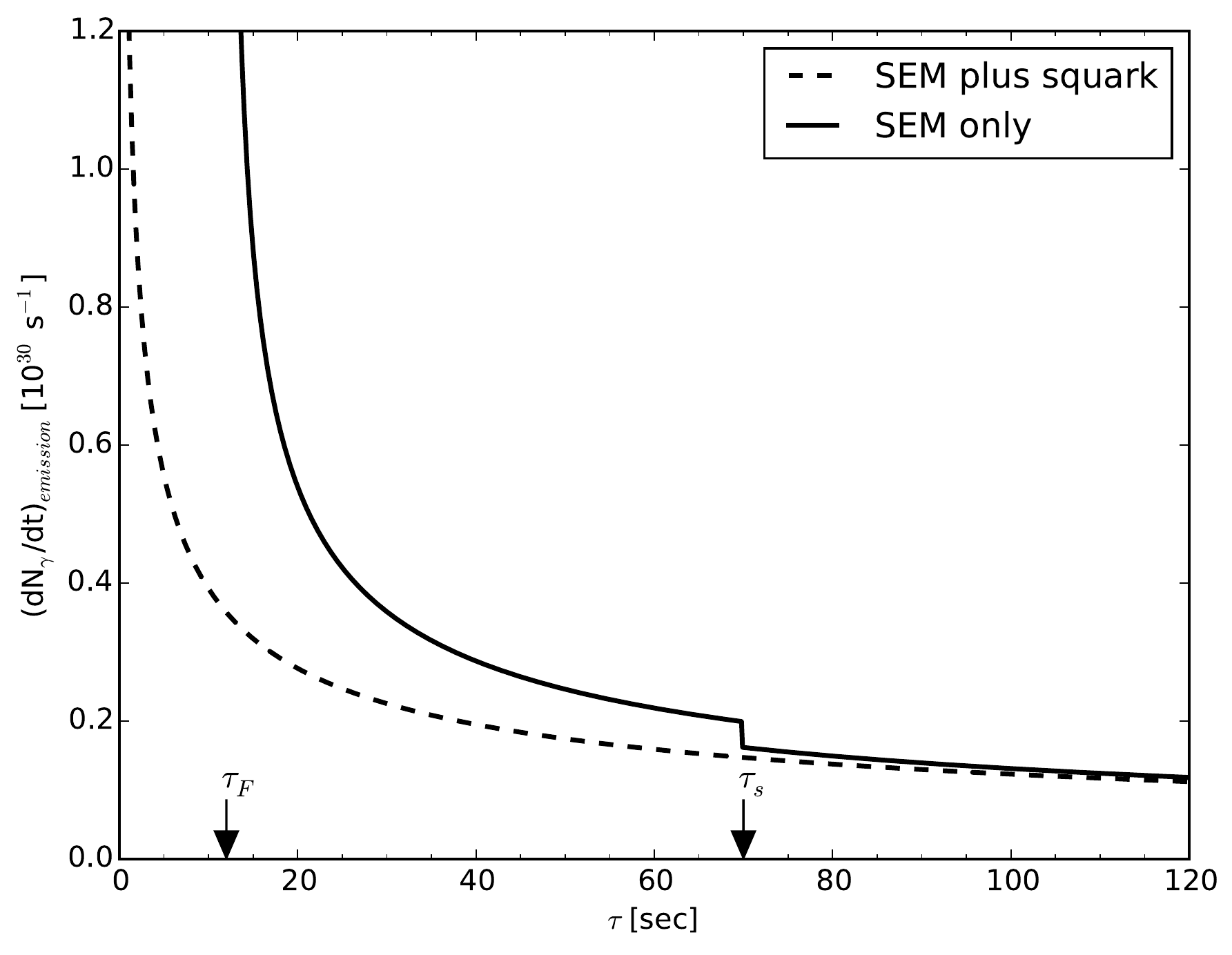}
\caption{Black hole photon emission rate of an SEM BH and a BSM (SUSY) BH as a function of the remaining lifetime of the SEM BH.
The time threshold for squark emission by the BSM BH is $\tau_{sq}= 70$ s;
the completion of the BSM BH burst occurs at $\tau_{F}= 12$ s; and the squark emission is
conservatively estimated to enhance the observable photon rate by $23\%$, as discussed in Section~\ref{subsec:susyphotons}.
Solid curve: the BSM black hole with squark radiation.
Dashed curve: the SEM black hole.
\label{fig:FIGY2}}
\end{center}
\end{figure}

\subsection{Statistical Estimate of Detection Sensitivity to Squark Emission} \label{subsec:susyphotons}

Let us make a rough estimate of the observational sensitivity to the SUSY squark threshold. As
shown in Fig.~\ref{fig:pbh_limits}, a likely search interval for the HAWC observatory lies within the range
of 10--100 s. The rest mass of a squark which would fall within this search window range is then given by 
\begin{equation}
m_{sq} c^2 \simeq 7.8 x_{p,s=0} \left(\frac{1\ \rm{s}}{\tau_{sq}}\right)^{1/3}\, \rm{TeV}
\end{equation} 
(see Eq.~\ref{eq:temptau}), i.e., $m_{sq}c^2\simeq$ 5--10 TeV for $10\ \rm{s}\lesssim \tau_{sq} \lesssim 100\ \rm{s}$. Because the value of $\alpha$ will increase due to the
Hawking radiation of the SUSY states, the actual remaining time will be somewhat shorter, and the squark
mass range somewhat higher than this estimate.

The observational signature of a SUSY superpartner threshold being passed by the BH would be an enhanced
photon rate at shorter remaining burst lifetimes (i.e., when $T_{BH}\gtrsim T_{sq}$), compared to the
rate at earlier times.  To estimate the extent to which discerning such a rate increase is feasible, we
consider the following conservative simplified model. A squark is expected to decay into a quark and the
Lightest SUSY Particle, typically a neutralino.  If the neutralino mass is not a large fraction of the initial
squark mass, then the quark should generate a photon spectrum similar to that generated by a directly Hawking-radiated
quark of the same initial energy as the squark.  To count the number of degrees of freedom of the initial
squark, we conservatively assume a squark of a single handedness and hence only 6
degrees of freedom (particle, antiparticle and 3 colors modes), rather than the 12 squarks degrees of freedom assumed in Eq.~\ref{eq:sqmassloss}. Recalling from Section~\ref{sec:Hawking_rad} that 
$\Psi_{s=0} = 2.8 \Psi_{s=1/2}$, we
expect that each squark degree of freedom is Hawking-radiated 2.8 more often than each quark degree of
freedom when $T_{BH}\gtrsim T_{sq}$. Thus, noting that there are $72$ total degrees of freedom for the $s=1/2$ SEM quarks, the
squark emission enhances the overall BH photon rate by a factor of about $6 \times 2.8 \div 72\simeq 23\%$. Fig.~\ref{fig:FIGY2}
compares the photon emission rate for the SEM BH and the BSM BH with $\tau_{sq} = 70$ s and $\tau_{F} = 12$ as a function of the
remaining lifetime of the SEM BH, assuming that both BHs had the same mass when $\tau \gg \tau_{sq}$ and the squark emission enhances the observable photon rate by $23\%$. 

In the same spirit, let us use a simple analysis to estimate the detectability of this photon rate increase. Let us assume that the power law of the BH photon burst emission time
profile shown in Fig.~\ref{fig:pbh_lightcurve} is little changed after the
SUSY superpartner threshold is reached, i.e., that the increase in $\alpha$ due to the SUSY states at $\tau_{sq}$ is negligible. We
compare the number of photons in the final $10$ seconds of the BH burst, with the number of photons seen in an
earlier longer interval, say between 80 and 200 seconds prior to the end of the burst. Assuming the photon rate
is enhanced by $\sim 20\%$ as estimated above for a squark threshold in the range
$10\ \rm{s} \lesssim \tau_{sq} \lesssim 80\ \rm{s}$, and applying the power law of
Fig.~\ref{fig:pbh_lightcurve}, the ratio of the number of photons in the 80-200 s interval is 0.97 times that
in the 0-10 s interval for the SEM BH, while it is 0.79 times that in the 0-10 s interval for the SUSY BH.
Taking the case considered in Fig.~\ref{fig:pbh_lightcurve_hawc} of a nearby PBH
burst at 0.015 parsec whose final 10 seconds would produce a photon count of $\sim 1400$ photons at the
HAWC observatory, we find, considering Poisson fluctuations only in the number of signal photons and
ignoring background fluctuations, that the SEM BH burst and a BSM BH burst with a squark threshold of $10\ \rm{s} \lesssim \tau_{sq} \lesssim 80\ \rm{s}$ would be
distinguishable with a significance above 4 standard deviations.

Thus it is clear that detection of a nearby PBH burst has an
interesting sensitivity to multi-TeV squark states which are currently inaccessible with hadron colliders.
Further refinements of these calculations and other SUSY models will be presented in a separate paper.

\section{Discussion}\label{sec:discussion}

Our analysis in Sections~\ref{sec:pbh_theory} - \ref{sec:pbh_search} assumes the SEM of black hole radiation based on the
Standard Model of high-energy physics and the Hawking radiation of fundamental particle species, such as
quarks and gluons, as initially asymptotically free particles.
If a direct PBH burst search is unsuccessful, the null result will constrain
the local density of such PBH burst events but any derived upper limit will depend on the validity of the SEM.
Alternative emission theories make different predictions for the PBH photon emission rate and/or spectrum. Here we discuss some issues arising from our SEM-based analysis.

\subsection{The Pion Fragmentation Function for Quarks and Gluons}\label{dis:pion_frag}

In this paper, we assumed that all Hawking-radiated quarks and gluons fragment
and hadronize into pions, according to the fragmentation function in Eq.~\ref{eq:hff}.
Is the pion fragmentation function employed in Section 2 a realistic approximation?

Fragmentation functions have been extracted from accelerator data, e.g. from
electron-positron collider experiments. The heuristic model in Eq.~\ref{eq:hff} agrees
qualitatively with the empirical fragmentation functions~\cite{deFlorian2007},
although the quantitative accuracy is limited.
Two aspects of the fragmentation process are absent from the heuristic model.
Firstly, the assumption that all flavors of quarks eventually fragment equally and completely
into pions is not strictly true. The fragmentation steps will also produce
other mesons and baryons. In addition, heavy particles such as the top quark, $W^\pm$, $Z^0$ and H initially decay into lighter quarks rather than directly
undergoing hadronic fragmentation into pions; the fragmentation function
into pions for these heavy particles will feature fewer pions at high $z$ than for light quarks, and somewhat
more pions with $z < 0.3$ or so.  Secondly, QCD fragmentation functions
depend to some degree on the energy scale of the process, whereas
Eq.~\ref{eq:hff} is scale-invariant.

An enhanced treatment of fragmentation and hadronization is possible
using either more detailed QCD fragmentation functions or Monte
Carlo simulations for the fragmentation and hadronization using parton
shower codes like Pythia~\citep{Pythiya2008} or Herwig~\citep{Herwig2013}. As we noted in Section~\ref{subsec:pionfrag}, though, the function
Eq.~\ref{eq:hff} implies an average energy of final state photons and a multiplicity of final state photons per initial
parton which match the $T_{BH}^{1/2}$ scaling of photon average energy and multiplicity found using a HERWIG-based Monte Carlo simulation for $1\ \rm{GeV} \leq T_{BH} \leq 100\ \rm{GeV}$ BH
emission spectra~\cite{MacGibbon1990}. We also showed in Section~\ref{subsec:time-int} that the time-integrated photon
spectrum derived using the fragmentation function of Eq.~\ref{eq:hff} is in good agreement with the parameterization
of the time-integrated spectrum derived by fitting the results of the HERWIG-based Monte Carlo BH simulations. Thus the
fragmentation function given in Eq.~\ref{eq:hff} is adequate for deriving a good estimation of the overall instantaneous photon BH emission spectra.

\subsection{Photospheres and Other Models of Intrinsic Interaction of BH Radiation}

The SEM assumes that relativistic quarks and gluons emitted as Hawking
radiation escape as asymptotically free particles from their creation region close
to the BH horizon (i.e., they do not undergo significant interactions with other Hawking-radiated particles
over distances at least up to of order $10^{-15}\ \rm{m}$ appropriately Lorentz-transformed),
analogous to quark and gluon jet creation in high-energy collisions in accelerator
experiments. (See \cite{MacGibbon2008} for the details of this analogy.)
Over distances of a few fermi appropriately Lorentz-transformed, the QCD quanta then
undergo fragmentation and hadronization,
consistent with observations of high energy accelerator collisions.

Some authors, however, have proposed that in the neighborhood of the BH the
radiated particles undergo additional interactions.
In these scenarios the Hawking radiation after emission self-interacts to form
a dense photosphere around the microscopic black hole~\cite{Heckler1997a,Heckler1997b,Kapusta1999}.
The emission rate is not modified but
the particle energies are degraded to lower energies in the
vicinity of the BH, resulting in a photon spectrum which, at high photon
energies, would be less than that predicted by the SEM (Fig.~\ref{fig:alphaSEM}). Although the limit derived
by the 100 MeV emission from a Galactic or cosmological distribution of PBHs would be only slightly modified,
the probability of detecting the high energy gamma ray or cosmic rays bursts from individual PBHs is
significantly weakened in photosphere models. For example, the Heckler photosphere model predicts 
that the photon flux above $E_{\gamma}\simeq 1$ TeV emitted by a $T_{BH} = 1$ TeV black hole over its
remaining lifetime is approximately 4 orders of magnitude less than the $E_{\gamma}\gtrsim 1$ TeV flux
predicted by the SEM analysis~\cite{Heckler1997b}. In the Heckler model, the high-energy time-integrated
photon spectrum decreases as $E_{\gamma}^{-4}$, not as $E_{\gamma}^{-3}$ as shown for the SEM case in Fig.~\ref{fig:photspect5DT}.

Recent detailed re-analysis of the published photosphere scenarios, however, has strongly argued
that the conditions required for the production of intrinsically-induced photospheres are not
met around Hawking-radiating black holes~\cite{MacGibbon2008}. In particular, the Hawking
flux emission rate implies that there is insufficient causal connection between the majority of consecutively
emitted particles for QED or QCD interactions to occur between them; the quantum conservation laws and
available energy per Hawking-radiated particle, together with the suppression of $\Gamma$ (see Section~\ref{sec:pbh_theory}) near rest mass
thresholds, prevent the formation of a QCD photosphere as the BH transitions through the QCD confinement
scale; and the long formation distance required for the production of any final state created in an
interaction prevents an individual particle undergoing multiple interactions close to the BH. 
Although one should be cognizant that intrinsically-induced photospheres
would change the observational characteristics and limits, we expect that the search
methods will be primarily based on the SEM for the next generation of PBH searches.

\subsection{Modification of BH Burst by Ambient Environment}

Although models for intrinsically-produced photospheres do not seem viable,
the possibility exists in the SEM that the observable signal may be modified if the PBH is
embedded in, for example, a region of ambient dense plasma or a strong
magnetic field~\cite{MacGibbon2008}.

Rees has proposed a model~\cite{Rees1977} in which the high
energy electrons and positions emitted in the final BH burst form
a relativistically-expanding conducting shell. The conducting shell is then braked by the ambient Galactic
magnetic field, generating a strong radio pulse. The original Rees model assumed that these were exclusively electrons and
positrons, and the remaining BH mass, were emitted in one instant once the BH temperature reached
$T_{BH}\simeq 0.16$ GeV. Re-analyzing the model using the SEM, extrapolating to higher $T_{BH}$ the emission
spectra of Ref.~\cite{MacGibbon1990} which incorporate quark and gluon emission, and taking a typical interstellar magnetic field
of strength $B\simeq 5\times 10^{-6}b$ G where $b=O(1)$, MacGibbon found that the conditions for the
generation of an electromagnetic pulse are not met until the BH
mass reaches $M_{pulse}\simeq b^{-1/2}(\alpha (M_{pulse}) / \alpha_{SM})^{1/2}$ g~\cite{MacGibbonCarr1991}. The electromagnetic pulse
would now be seen at about $E_{\gamma}\simeq 70 b^{3/2}(\alpha (M_{pulse}) / \alpha_{SM})^{-1/2}$ TeV
(with a duration $\Delta t \lesssim 10^{-25}$ s much less than the time resolution of any detector) and
contain a total energy of $E_{pulse}\simeq 10^{25} b^{-1/2}(\alpha (M_{pulse}) / \alpha_{SM})^{-1/2}$ GeV.
If the conditions for the pulse are met, there would be $N_{\gamma} = E_{pulse}/E_{\gamma}\simeq 10^{23}b^{-2}$
photons emitted in the pulse. For a typical interstellar magnetic field ($b=O(1)$), the pulse would
thus be much weaker than the PBH lightcurves of Figs.~\ref{fig:pbh_lightcurve} and
~\ref{fig:pbh_multi_lightcurve}. Although these estimates should not be regarded as
precise, because they involve the extrapolation of the BH emission spectra to energies well above accelerator
energies, it can be stated that the electromagnetic
pulse generated by a bursting SEM BH in the Galactic magnetic field would have a wavelength in the gamma-ray
range, not radio range as in the original Rees analysis, and occur at a much smaller $M_{BH}$ thus producing a less-detectable signal.

\subsection{PBH Evaporation Events in the Hagedorn Model}

Some previous studies of BH burst emission have assumed
the Hagedorn model, also called the statistical bootstrap model.
This particle physics model, which arose before the existence of quarks
and gluons was confirmed in terrestrial accelerators, postulates that there is an
exponentially rising spectrum of meson resonances once a threshold temperature
$\Lambda_{QCD}$ has been reached.

The Hagedorn PBH model assumes each of the meson resonances is an independent degree
of freedom of Hawking radiation. Thus in the Hagedorn model, the function $\alpha (M_{BH})$
exponentially increases in this temperature regime and the PBH luminosity is
correspondingly enhanced. The model also assumes that the remaining mass of the BH will
be emitted around this temperature producing a stronger final burst that will be confined
to lower photon energies ($\lesssim 1 $ GeV), in contrast to the SEM burst.

However, with the discovery of quark and gluon jets in accelerator collisions above
$\Lambda_{QCD}$, the Hagedorn model is no longer a viable description of particle production
in such collider events. Moreover, detailed consideration~\cite{MacGibbon2008} of the particle separation,
energies and timescales of Hawking radiation of QCD particles indicates that the asymptotic freedom of QCD which describes jet
production at hadron colliders applies to Hawking radiation\footnote{The analysis of Ref.~\cite{MacGibbon2008} also shows that the conditions for the production of quark-gluon
plasma are not met around the black hole.}. Additionally, because the Hawking radiation of a particle correspondingly
reduces the BH mass and hence increases the BH temperature, any Hagedorn phase can at most
be temporary with the BH transitioning quickly to temperatures above the
Hagedorn regime. The form of the absorption factor $\Gamma$ of Eq.~\ref{eq:hawking}
also strongly suppresses the Hawking emission of a species when $T_{BH}$ is close
to the rest mass threshold of the species, thus weakening the signal from any temporary Hagedorn
regime. Taken together, these considerations strongly argue against the Hagedorn model applying to BH
emission or producing an enhanced PBH burst signal at the detector.

\subsection{Do Very Short Gamma-ray Bursts Originate from PBHs?}

Studies of very short gamma-ray bursts (VSGRB),
with time durations $\lesssim 0.1\ \rm{s}$, tentatively suggest that these
events may form a distinct class of GRBs~\cite{Cline2005, Ukwatta2015grb}. The data used in these
studies come from BATSE, Fermi GBM, Swift, KONUS, and other
keV/MeV gamma-ray detectors~\cite{Cline2005,Cline2011,Czerny2011}. Evidence that the VSGRBs are distinctly different
from GRBs of longer duration includes the anisotropy on the sky of the
distribution of VSGRBs~\cite{Ukwatta2015grb} and the hardness of their
photon spectra. The VSGRB sky positions may be clustered close
to the anti-galactic center region~\cite{Ukwatta2015grb}, unlike short or long GRBs,
possibly indicating that the VSGRBs may have local origin.

These characteristics of VSGRBs have led to speculation that some fraction
of these events may be PBH bursts~\cite{Cline2005,Cline2011,Czerny2011}.
The highest photon
energies observed in these VSGRBs are less than 10 MeV.
The authors in Refs.~\cite{Cline2005,Cline2011,Czerny2011} compared the observed light curves with the total flux that would be seen from a nearby PBH burst, assuming that
the remaining mass of the PBH is converted into photons with energy well below $1$ MeV. The authors found
reasonable agreement between the shape of the observed and predicted time profiles, but achieved the best matches by assuming that BH emission is enhanced
as $T_{BH}$ approaches a phase transition or as conditions for a `fireball' photosphere set in. We
note, however, that the photon energies of these detectors are much lower than
the 50 GeV -- 100 TeV range which we have analyzed in this paper. Furthermore the behaviour of the quark and
gluon fragmentation and hadronization functions at photon energies well below $100$ MeV is uncertain and
our results can not be simply extrapolated to such low photon energies.
Although it is possible that the VSGRBs may ultimately be explained by
astronomical processes involving compact objects, such as neutron star mergers
in the Milky Way galaxy, we intend in a following paper to address the modelling of the BH burst gamma-ray
spectra below $1$ GeV using the SEM, motivated by the VSGRB observations and proposals for future telescopes in these lower wavelengths. In any case,
PBH searches at TeV-scale observatories should be based on the SEM at $T_{\rm BH} \gg 100$ MeV.

\section{Summary and Conclusions}\label{sec:conclusion}

In this paper, we have reviewed and analyzed the theoretical framework of the standard
BH emission mechanism and explored observational characteristics of the final burst. Moreover we have also explored and compared PBH burst search methods and differences between a PBH
burst and standard cosmological GRBs. Here are the main finding and conclusions of this
paper:

\begin{enumerate}
 \item We have developed improved approximate analytical formulae for the instantaneous
BH photon spectrum which includes both the directly Hawking-radiated photons and the photons resulting from the decay or fragmentation and hadronization of other directly Hawking-radiated species. Our analysis incorporates the most recent LHC Standard Model results.

 \item For the first time, we have calculated the PBH burst light curve (time profile) and studied its energy dependence both at the source and at the detector.

 \item At relatively low energies ($E_{\gamma} < 10$ TeV), the PBH burst light curve time profile does not show much variation with energy and is well described as a function of remaining burst lifetime by a power law of index $\sim -0.7$. However, at high energies, the PBH burst light curve profile displays significant variation with energy
that may be used as an unique signature of PBH bursts. In addition,
at high energies, the light curve profile shows an inflection region around 0.1 seconds.
The HAWC observatory is sensitive in this energy range for a sufficiently nearby PBH
and the above features in the
light curve may be used to uniquely identify PBH bursts.

  \item Compared to the burst Simple Search (photon counting) method, we have found that Maximum Likelihood Search methods using the PBH burst light curve are about 30\% more sensitive to PBH bursts. However, we also found that there is not a significant difference in sensitivity between 10-Bin Maximum Likelihood Searches and Unbinned Maximum Likelihood Searches.

  \item We have discussed in Section~\ref{sec:grb_vs_pbh} the expected differences
between PBH bursts and standard cosmological GRBs, in particular that the PBH bursts evolve higher into the TeV band and are typically not expected to be accompanied by an afterglow.

  \item We have shown in Section~\ref{sec:BSM} the possibility that evidence of Beyond the Standard Model physics may be discernible in the detector signature of a PBH burst. In particular, a squark state with a rest mass of 5-10 TeV is expected to produce a detectable change in the time dependence of the arrival rate of TeV photons emitted in the final 200 seconds of the PBH burst.

\end{enumerate}

\section{Acknowledgements}\label{Acknowledgements}

This work was supported by grants from the National Science Foundation
(MSU) and Department of Energy (LANL). TNU acknowledges the partial
support of this work by the Laboratory Directed Research \& Development
(LDRD) program at LANL. We would also like to thank
Wade Fisher of MSU for useful conversations on the likelihood fits,
Jing-Ya Zhu of MSU for discussions on current models of supersymmetry
and Sekhar Chivukula of MSU for useful conversations on Higgs field degrees of freedom and Extra Dimension models. We also thank the anonymous referee
for comments that significantly improved the paper.

\bibliographystyle{ieeetr}
\bibliography{my_references}

\end{document}